\documentclass[11pt]{article}
\usepackage[verbose,tmargin=2.4cm,bmargin=2.4cm,lmargin=2.4cm,rmargin=2.4cm]{geometry}
\usepackage[english]{babel}
\usepackage[utf8]{inputenc}
\usepackage{graphicx}
\usepackage{setspace}
\usepackage{parskip}
\usepackage{float}
\usepackage{booktabs}
\usepackage{mathrsfs}

\usepackage{commath}
\usepackage{lmodern}
\usepackage{framed}

\usepackage{titling}       
\usepackage{ragged2e}
\usepackage[labelfont=bf]{caption}
\usepackage{subcaption}
\usepackage{siunitx}
\usepackage{capt-of}
\usepackage[most]{tcolorbox}

\usepackage{amsmath,amsfonts,amssymb,amsthm}
\usepackage{mathtools}
\usepackage{physics}

\usepackage{dsfont}
\usepackage[linesnumbered,ruled]{algorithm2e}
\usepackage{scalerel}
\usepackage{tikz}
\usetikzlibrary{automata,arrows,positioning,calc}
\usepackage{xcolor} 
\usepackage{epstopdf}
\usepackage{placeins}
\usepackage{verbatim}

\usepackage{multirow}

\usepackage{url}
\usepackage[pdfstartview=FitV, urlcolor=blue]{hyperref}
\usepackage{authblk}

\usepackage[title]{appendix}

\usepackage[sort, numbers]{natbib}

\newcommand{\y}{\mathbf{y}}
\newcommand{\bc}{\mathbf{c}}
\newcommand{\bchat}{\mathbf{\hat{c}}}

\newcommand{\J}{\mathcal{J}}
\newcommand{\E}{\mathbb{E}}
\renewcommand{\Pr}{\mathbb{P}}

\DeclareMathOperator*{\argmax}{arg\,max}

\newcommand{\subS}{{}_{{\scalebox{0.5}{$\mathrm{S}$}}}}
\newcommand{\subI}{{}_{{\scalebox{0.5}{$\mathrm{I}$}}}}
\newcommand{\subR}{{}_{{\scalebox{0.5}{$\mathrm{R}$}}}}
\newcommand{\subD}{{}_{{\scalebox{0.5}{$\mathrm{D}$}}}}

\newcommand{\subSo}{{}_{{\scalebox{0.5}{$\mathrm{S,\!o}$}}}}
\newcommand{\subSr}{{}_{{\scalebox{0.5}{$\mathrm{S,\!r}$}}}}
\newcommand{\subSf}{{}_{{\scalebox{0.5}{$\mathrm{S,\!f}$}}}}
\newcommand{\subSj}{{}_{{\scalebox{0.5}{$\mathrm{S,\!j}$}}}}
\newcommand{\subSk}{{}_{{\scalebox{0.5}{$\mathrm{S,\!k}$}}}}
\newcommand{\subIo}{{}_{{\scalebox{0.5}{$\mathrm{I,\!o}$}}}}

\newcommand{\subIj}{{}_{{\scalebox{0.5}{$\mathrm{I,\!j}$}}}}
\newcommand{\subIk}{{}_{{\scalebox{0.5}{$\mathrm{I,\!k}$}}}}

\newcommand{\subRj}{{}_{{\scalebox{0.5}{$\mathrm{R,\!j}$}}}}
\newcommand{\subRk}{{}_{{\scalebox{0.5}{$\mathrm{R,\!k}$}}}}

\newcommand{\cS}{c\subS^{}}
\newcommand{\cSS}{c\subS^{*}}
\newcommand{\cSoS}{c\subSo^*}
\newcommand{\cSj}{c\subSj^{}}
\newcommand{\hatcS}{\hat{c}\subS^{}}
\newcommand{\hatcSS}{\hat{c}\subS^{*}}
\newcommand{\cI}{c\subI^{}}
\newcommand{\cR}{c\subR^{}}
\newcommand{\cSmyo}{c\subS^{{\scalebox{0.5}{\#}}}}

\newcommand{\cSbar}{\bar{c}\subS^{}}
\newcommand{\cIbar}{\bar{c}\subI^{}}
\newcommand{\cRbar}{\bar{c}\subR^{}}

\newcommand{\uS}{u\subS^{}}
\newcommand{\uI}{u\subI^{}}
\newcommand{\uR}{u\subR^{}}

\newcommand{\uSbar}{\bar{u}\subS^{}}
\newcommand{\uIbar}{\bar{u}\subI^{}}
\newcommand{\uRbar}{\bar{u}\subR^{}}

\newcommand{\VS}{V_{{\scalebox{0.5}{$\mathrm{S}$}}}}
\newcommand{\VI}{V_{{\scalebox{0.5}{$\mathrm{I}$}}}}
\newcommand{\VR}{V_{{\scalebox{0.5}{$\mathrm{R}$}}}}
\newcommand{\VD}{\Psi\subD}

\newcommand{\nt}{r}

\newcommand{\newest}{\textcolor{black}}

\newtheorem*{prop}{Proposition}
\newtheorem*{remark}{Remark}

\newif\ifinlinefigures

\inlinefigurestrue

\newif\ifmathlineno

\mathlinenofalse

\title{\LARGE \bf
Behavioral patterns and mean-field games in epidemiological models
}
\author[1]{Finnegan Buckley}
\author[2]{Alexander Vladimirsky}
\affil[1]{Department of Mathematics, Cornell University, Ithaca, NY, {\tt ftb24@cornell.edu}}
\affil[2]{Department of Mathematics and Center for Applied Mathematics, Cornell University, Ithaca, NY, {\tt vladimirsky@cornell.edu}}

\begin{document}

\maketitle

\spacing{1.7} 
\begin{abstract}
We introduce a new type of Mean Field Game epidemiological models, in which 
subpopulations have different behavioral patterns:
some people exhibit a type of ``rational'' self interested behavior by choosing Nash-equilibrium long-term strategies 
while others follow pre-specified ``non-rational'' patterns (e.g., either sticking to their usual habits or trying to mimic those around them).  Our new B-MFG models allow for occasional behavioral switches, which rational individuals also take into account when formulating their Nash-equilibrium strategies. Although this modeling approach is general, here we develop it for individuals choosing their ``contact rates'' within a particular Susceptible-Infected-Recovered-Susceptible-Dead  (SIRSD) epidemic model.  
The latter is based on  a frequency-based force of infection with proportionate mixing 
and the mortality rate that rapidly increases once the proportion of infected individuals exceeds some prescribed threshold, resulting in a strain on medical resources.
Numerical tests illustrate the properties of our model and highlight the ways in which additional/non-rational behavioral patterns and behavioral switching increase the impact of infectious diseases.  The paper aims to build a bridge between two distinct communities of epidemiological modelers and to promote the consideration of behavioral patterns in broader Mean Field Games literature.
\end{abstract}

\section{Introduction}
\label{s:intro}
Mathematical epidemiology is a classical area of applied mathematics, 
which saw a significant increase in research activities since the onset of COVID pandemic.
Many mathematicians seized the opportunity to bring new tools and ideas to these problems of obvious and immediate societal importance.  
Among them, the experts in Mean Field Games (MFGs) have proposed a number of models to explain individuals' choices on how much to follow official rules and recommendations (on vaccinations, social distancing, mask wearing, etc)
\cite{reluga2010game, tembine2020covid, doncel2022mean, aurell2022optimal, aurell2022finite, olmez2022modeling, roy2023recent, bremaud2024mfg, bremaud2025mean, schnyder2025understanding, liu2025incorporating, doebeli2026partial}.
In MFG literature,
individuals are assumed to be strategic planners.
Their choices are viewed as 
fully rational in a selfish or decentralized manner: ``optimal'' from
the point of view of individual decision makers, but often quite far from ``socially-optimal''.  While obviously an important contribution, such models do not account for the fact that people are not uniform in their priorities, rationality, analytic abilities, consistency in carrying out their preliminary plans, and self-awareness.  
Our goal in this paper is to develop extended MFG-type models that address some of these limitations, and to explore the effect of such features in the context of epidemiology.

The importance of population heterogeneity and behavioral patterns is widely recognized in modeling the spread of infectious diseases.  
Many of the classical models already split population into multiple groups based on their frailty and propensity to spread the infection.  
E.g., the need to consider the role of super-spreader {\em core groups} separately
is a particularly common feature in modeling sexually-transmitted diseases \cite{hethcote2014gonorrhea}.
But more recent 
behavior-epidemiology
models go much further, allowing for individuals' occasional switches in behavioral patterns, and modeling how the frequency of such switches is affected by individuals' fatigue from following the rules, peer pressure, and changing 
(real or perceived) risk of infection;
see, for example,
\cite{pant2024mathematical, lejeune2024mathematical, lejeune2025formulating, oveson2025modeling, CHEN2025109345, nguyen2025complex}.
Endogenous rates of behavioral switching are sometimes also derived 
using models of {\em social learning}, most often in the framework
of Evolutionary Game Theory (EGT) \cite{manfredi2013modeling, glaubitz2020oscillatory, martcheva2021effects, agusto2022isolate, saadroy2023dynamics, agusto2026algorithms}.
However, in all these cases, contact rates and other parameters within each behavioral pattern are defined a priori --  either as pre-specified constants or as pre-specified functional forms (e.g., as a function of the current disease prevalence or hospitalization level \cite{oveson2025modeling}), with no attempt to explain why each subpopulation adopts their specific set of parameter values or functional forms. 
Such models can be calibrated using historical data, but this sometimes makes it harder to use them in the context of public policy; e.g., in predicting how people's behavior (and the resulting epidemiological trajectory) will change in response to proposed guidance updates when the relevant historical data is sparse or unavailable. 

Game-theoretic models based on strategic planning are, in many ways, the opposite of the above phenomenological approach, with individual behavior typically explained through the lens of trying to maximize one's personal ``rewards'' or ``returns''.  
When many individuals (or ``players'') make their decisions independently while affecting each other, the concept of optimality has to be replaced by the notion of Nash equilibrium:  
a combination of individual decisions/strategies such that no individual can improve their personal returns by unilaterally changing their own strategy.
But computing and analyzing Nash equilibria is generally harder in dynamic 
games involving a large  
number of players.  Mean Field Games (MFGs) were introduced to address the latter challenge \cite{lasry2006jeux_1, lasry2006jeux_2, huang2006large} and have by now found much success, particularly in economics and engineering applications.
The key idea in MFGs is to consider a game with {\em infinitely many}  
``small'' players, each of whom 
interacts with an evolving density of players rather than with specific individuals\footnote{
MFGs are certainly ``not the only game in town.'' While our main focus is on extending the MFGs,
we later provide 
a detailed comparison of them with two other game-theoretic modeling approaches 
in Appendix~\ref{appendix:other_games}.
But to avoid a possible terminological confusion,
we note explicitly already here that MFGs are very different from ``games against the field,'' which were introduced much earlier by John Maynard Smith as a part of the Evolutionary Game Theory (EGT) 
\cite{smith1982evolution}.
In contrast with EGT models, in MFGs
(1) each player has a state (quite distinct from their strategy);
(2) the game is dynamic (lasting over a long or infinite horizon), with each player's state changing as a result of their personal strategy, the evolving distribution of all players' states and, in some cases, of strategies adopted by all others;
and (3) the strategies are not inherited or learned from others; instead, they are chosen rationally (in a time-dependent way) to minimize the cost or maximize the returns accumulated over time.  In MFGs, what a player chooses to do now depends on  predicted consequences; in EGT, all changes to the composition of strategies are driven by the current composition and the current strategy ``payoffs'' only.

The assumptions that individuals are long-term planners and recognize that their collective actions have consequences for the course of epidemic are also fundamental for MFGs. The goal of predicting those consequences before making personal decisions significantly differentiates MFGs from other game-theoretic models, in which individuals are assumed to lack predictive abilities (e.g., \cite{fenichel2011adaptive, fenichel2013economic, morin2013sir, perrings2014merging, espinoza2022heterogeneous, espinoza2024adaptive, espinoza2025impact})
and thus use a very different notion of ``optimal behavior''.
}.  
Computationally, this leads to a coupled system of 
differential equations describing the evolution of players' density and players' optimal (Nash equilibrium) behavior.  For MFGs over a finite state space, this yields a two point boundary value problem for a system of Ordinary Differential Equations (ODEs)
\cite{gomes2013continuous}.  This is a natural fit for epidemiological models, 
where possible individual states are based on their current health status (e.g., Susceptible, Infected, Recovered, or Dead -- giving rise to ``SIRSD compartment models'') and where the ODEs typically arise as a large population limit of interacting Markov chains. 

The rational optimization perspective allows for comparing the epidemiological and economic impact of Nash equilibrium solutions 
(when individuals make independent decisions) 
with what could be achieved with  
``socially optimal'' solutions (if everyone followed the optimally formulated official guidance, recovered in the framework of Mean Field Control) \cite{cho2020, roy2023recent}.
Other models go beyond this comparison and formulate the problem as a ``Stackelberg game'' between the authorities, who choose a public policy under assumption that people will respond to them in a rational but selfish manner, and individuals, who react to that known public policy by pursuing Nash equilibrium personal strategies \cite{aurell2022optimal}.  
While the above are usually formulated with the assumption that the logic of all
participants is homogeneous, there is also a broader literature on {\em multipopulation MFGs} \cite{cirant2015multi,bensoussan2018mean,Dayanikli2024}, which allows each subpopulation to have different optimization goals and dynamics.  Such features are obviously useful in epidemiological context; e.g., for elderly/frail individuals, the rate of death while infected is usually higher and the frequency of interactions between different age groups is often non-uniform as well \cite{aurell2022finite};
an individual's cost of following the official guidelines \newest{might} also be correlated with their socio-economic status \cite{liu2025incorporating}.  

But the full rationality and consistency assumptions remain fundamental even in such standard 
multipopulation MFG
models, with all individuals assumed to have the full knowledge of 
others' priorities and risks, ability to compute a personal (time-dependent) Nash equilibrium strategy, also predicting how similar optimization carried out by others will influence the 
epidemiological situation long term,
and consistency in implementing the selected Nash strategy throughout the course of epidemic.
This idealized description is hardly fitting for most 
humans,
and the resulting discrepancies can have significant consequences if
a Stackelberg-optimal public policy is adopted in practice.

In this paper, we suggest an approach for relaxing several of these assumptions by recognizing that only some of the people are highly rational 
(in the above technical sense), while others frequently adopt behavioral patterns that are not grounded in game theory.
More specifically, our new {\em MFG model with behavioral patterns} (which we will refer 
to as {\em B-MFG}) includes
three subpopulations: Reckless/stubborn individuals, who are largely risk-indifferent and
ignore the epidemiological situation; Followers/consensus-seekers, who try to adopt the ``population-averaged'' behavior; and rational Optimizers, who make their decisions strategically in an MFG-fashion, taking the behavior of all subpopulations into account\footnote{
The terms ``rational Optimizers'', ``Reckless'', and ``Followers'' are inherently imprecise and lack standard definitions.
We merely use them
as convenient shorthand names for specific behavioral patterns defined rigorously in \S~\ref{s:MFG_models}.}.  
Because many people are inconsistent in carrying out their long-term plans, in our model we also allow for individual's occasional random switching among these behavioral patterns, with switching rates possibly dependent on the epidemiological situation or fatigue from restricting their contacts.

Since one of the goals of this paper is to foster a dialogue between two distinct research communities, this also influences our exposition: we primarily focus on fundamental modeling assumptions and their consequences illustrated through numerical 
experiments, leaving other important aspects (theoretical, numerical, and data-centric) for the future.
To make the discussion concrete, we introduce a specific simple SIRSD ODE model in \S~\ref{ss:baseline_model}.  We then develop its generalized MFG version in \S~\ref{s:MFG_models}: starting with the optimization task faced by a single rational individual when others' strategies are already known, deriving a Nash description when all such individuals make decisions simultaneously, then adding other behavioral subpopulations and the possibility of switching between them.  The numerical results based on this general model are presented in \S~\ref{s:results}.  The limitations and directions for future work are discussed in \S~\ref{s:conclusions}.  
In Appendix~\ref{appendix:other_games}, we compare the fundamental principles of  MFGs/B-MFGs 
with two other game-theoretic approaches. 
In Appendix~\ref{appendix:SIRS_equilibria}, we also include the study of equilibria in 
the baseline model if mortality is omitted.

\subsection{The baseline SIRSD model.}
\label{ss:baseline_model}

We assume that at any given time each individual can be either
Susceptible (S), Infected (I), Recovered/Immune (R) or Dead (D), with the corresponding letters also used to encode the fraction of individuals in each state.
For the sake of simplicity, we also ignore the births and disease-unrelated deaths.
We further assume that
\begin{itemize}
\item 
Susceptible, Infected, and Recovered individuals each on average 
contact $\cS, \cI,$ and $\cR$ other individuals (respectively) per unit time.
For now, these will be treated as constants, but starting in the next section we will interpret these as individual's time-dependent control variables.
\item 
$\beta$ is the probability of infectious disease transmission for each contact of
a Susceptible with any Infected individual. 
\item 
The overall probability of a Susceptible becoming Infected per contact is thus

\begin{equation}
\label{eq:original_p}
p \; = \;
\underbrace{\beta}_{\Pr(\text{infection upon contacting an infected})} \, \overbrace{\frac{\cI I}{\cS S \, + \, \cI I \, + \, \cR R}}^{\text{Fraction of contacts made by Infected}}
\end{equation}
This formula is consistent with a frequency-based force of infection and
the so-called {\em proportionate mixing} assumption \cite{nold1980heterogeneity, HETHCOTE198785} with health-status-dependent contact rates.
Models built using a frequency-based force of infection are increasingly found to be 
appropriate for a broad range of infectious diseases; see, for example \cite[Table 1]{ferrari2011pathogens}.
\item
$\frac{1}{\mu}$ is the average time one stays infected until recovery; we thus refer to $\mu$ as the recovery rate.
\item
As the immunity starts to wane, $\frac{1}{\gamma}$ is the average time one stays immune after recovery;
we thus refer to $\gamma$ as the immunity waning rate.
\item 
$\delta(I) $ is the (monotone non-decreasing) per capita death rate of Infected due to the disease.
We use this dependence on $I$ to model the possible strain on medical services and infrastructure due to high infection levels.
(Thus, our model can also be described as ``SIRS with disease-induced and prevalence-dependent mortality''.)
While our general framework does not depend on a specific $\delta(I),$
in all of the following experiments we use a cubic spline-based death rate to model a \newest{rapid increase} from $\delta_1 \geq 0$ 
(when hospitals are operating normally) to $\delta_2 \geq \delta_1$ (when hospitals are overwhelmed by Infected):
\begin{equation}
\label{eq:death_rate}
\delta(I) \; = \; \left\{
	\begin{array}{ll}
		\delta_1,  & \mbox{if } I \le d_1; \\
		\delta_1+(\delta_2-\delta_1)\left(\left(\frac{I - d_1}{d_2-d_1}\right)^2\left(3-2\left(\frac{I - d_1}{d_2-d_1}\right)\right)\right), & \mbox{if } d_1\le I< d_2;\\
		\delta_2, & \mbox{if } I \ge d_2.
	\end{array}
	\right.    
\end{equation}
\end{itemize}

Using this notation and assumptions, the progression of epidemics can be found by solving a system 
\begin{align}
\label{eq:SIRSD_ODEs_original_S}
S' &\; = \; \overbrace{- p \cS S}^{\text{new infections}} \, + \, \overbrace{\gamma R}^{\text{waning immunity}}\\[1.1em]
\label{eq:SIRSD_ODEs_original_I}
I' &\; = \;  \overbrace{p \cS S}^{\text{new infections}} \, - \, \overbrace{\mu I}^{\text{recoveries}} \, - \, \overbrace{\delta(I) I}^{\text{new deaths}}\\[1.1em]
\label{eq:SIRSD_ODEs_original_R}
R' &\; = \;  \overbrace{\mu I}^{\text{recoveries}} \, - \, \overbrace{\gamma R}^{\text{waning immunity}}\\[1.1em]
\label{eq:SIRSD_ODEs_original_D}
D' &\; = \;  \overbrace{\delta(I) I}^{\text{new deaths}}
\end{align}
for $t \in (0,T]$, starting from the initial conditions 
$S(0) = S_0, \, I(0) = I_0, \, R(0) = R_0, \, D(0) = D_0.$

The table \ref{tab:params} summarizes our baseline parameter values and initial conditions.  These are used throughout the paper 
except where different values are specified explicitly.

\begin{table}[ht]
    \centering
    \begin{tabular}{|c|c|}
    \hline
         Initial Condition &  Value\\
         \hline
         
         $S_0$ & 0.995\\
         $I_0$ & 0.005\\
         $R_0$ & 0\\
         $D_0$ & 0\\
         \hline
    \end{tabular}
    \hfill
    \begin{tabular}{|c|c|c|}
         \hline
         Parameter &  Description & Value\\
         \hline
         $\gamma$ & Rate of loss of immunity & 1/90\\
         $\mu$ & Rate of recovery & 0.1\\
         $\beta$ & Chance of infection upon contact with infected& 0.15\\
         \hline
    \end{tabular}
    \hfill
    \par\bigskip
    \begin{tabular}{|c|c|c|}
         \hline
         Parameter & Description & Value\\
         \hline
         $\delta_1$ & Low mortality rate (when $I<d_1$)& 0.001\\
         $\delta_2$ & High mortality rate (when $I>d_2$)& 0.002\\
         $d_1$ & Maximum $I$ for the low mortality regime&0.15\\
         $d_2$ & Minimum $I$ for the high mortality regime&0.25\\
         \newest{$\VD$} & The death penalty & $-100$\\
         \hline
         
    \end{tabular}
    \caption{Baseline values for the initial conditions and ODE parameters.  
    }
    \label{tab:params}
\end{table}

\begin{figure}[hbt]
    \centering
    \includegraphics[width=11.9cm]{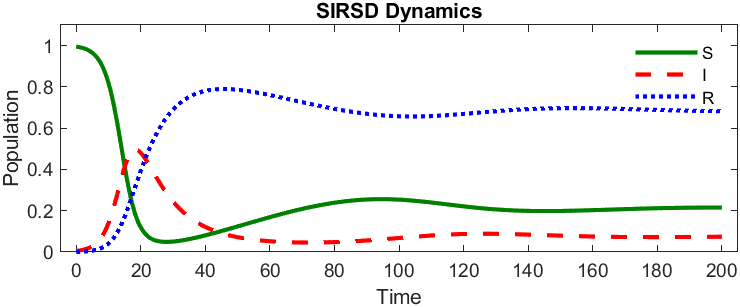}
    \caption{SIRSD epidemic dynamics over a 200 day time interval}
    \label{fig:noopt200}
\end{figure}

An epidemic trajectory resulting from this model is illustrated in Figure \ref{fig:noopt200} using the baseline contact rates $(\cS = 5, \cI = 3, \cR = 5),$ which will be motivated in the next section, starting from the initial conditions $D_0 = R_0 = 0, \, I_0 = 5 \times 10^{-3}, \, S_0 = 1- I_0,$ and computing until $T=200.$ 
This Figure tells an SIRSD-typical story with an initial growth in Infecteds reaching the maximum by $t \approx 18.4$, leading to an increase in the number of Recovereds up to $t \approx 45.6$; then the waning immunity leading to another local maximum in the number of Susceptibles by $t \approx 94.5$ and a subsequent second (much smaller) increase in the number of Infecteds around $t \approx 120.$  
To avoid cluttering the figures, we do not plot $D(t)$ but instead simply
report the total number of dead for each example. Since the death rates are relatively low, in this case $D(T) \approx 0.0286$.

\vspace*{-2mm}

While our focus is on model behavior over a finite horizon,
we note that with disease-induced mortality  
this model does not include an endemic equilibrium.
It is easy to see that $(S, 0, 0, 1-S)$ is an equilibrium for every $S \in [0,1]$
and there are no other equilibria as long as $\gamma, \delta_{\newest{1}}>0.$
For the sake of completeness, in Appendix \ref{appendix:SIRS_equilibria}
we also analyze the existence and stability of the endemic 
equilibrium when mortality is absent (i.e., with $\delta_1 = \delta_2 = 0$).

\section{MFG models and behavioral patterns}
\label{s:MFG_models}
Since this manuscript is intended to be of interest to a broader audience of epidemiological modelers,
we describe our approach in greater detail than it is usually done in MFG literature, focusing on modeling assumptions as well as on their mathematical and practical implications.  

Our focus will be on a mixed population of agents, all of whom change their personal contact rate independently
with the full knowledge of current epidemiological situation.
Some of these agents (referred to as ``Optimizers'') are assumed to be highly rational in a sense that they  
\begin{itemize}
    \item 
    strive to maximize their personal cumulative reward, which is computed by integrating over the time-horizon $[0, T ]$ their {\em instantaneous utility} function $u$ that changes based on their current status (S, I, or R) and the chosen contact rate;
    \item 
    can predict the future epidemiological trajectory if everyone's current and future contact rates are known;
    \item 
    can find Nash-equilibrium strategies, reflecting their understanding that all other Optimizers are simultaneously performing their own selfish maximization as well.
\end{itemize}
The logic of other agents (Reckless or Followers) will be mathematically described later, but we emphasize that 
they are not viewed as ``rational'' in the same sense as Optimizers.

We start by focusing on Optimizers: describing their instantaneous utility functions (\S~\ref{ss:utility}),
formulating the problem faced by a single Optimizer when everybody else's contact rates are known/fixed
(\S~\ref{ss:optimal_control}), and explaining how the Nash-equilibrium strategies are obtained in the MFG setting
when the entire population consists of Optimizers (\S~\ref{ss:Nash}).
We then finally consider a full B-MFG model with a mixed population (\S~\ref{ss:multi_pop_MFG}), in which the Optimizers take into account both the impact of non-optimizing agents {\em and} everybody's (including their own) propensity for occasional switches in behavioral patterns.

\subsection{Utility of Contact Rates}
\label{ss:utility}
Individuals need to make contact with others for economic gains and for their personal well being. Thus, we assume that an individual's instantaneous utility is a function of their contact rate with others.  
While our general model does not require a specific type of utility functions,
all of the numerical experiments in \S~\ref{s:results} are based on
the utility functions 
introduced in \cite{fenichel2011adaptive}.

\begin{figure}[hbt]
\centering
\includegraphics[width=10cm]{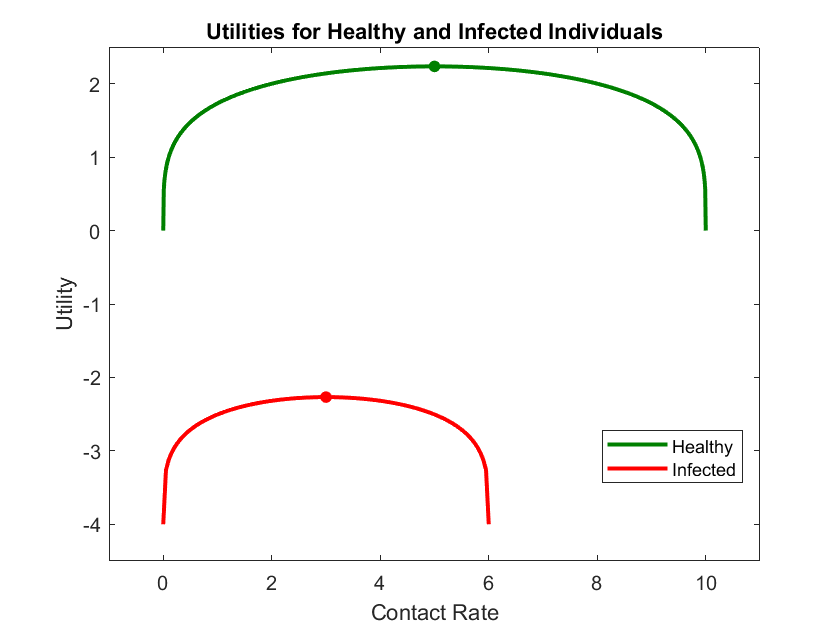}
\caption{ 
\label{fig:utilities}
Graphs of instantaneous utility functions for a healthy (susceptible or recovered) individual and an infected individual. The maximum values of the two functions are noted by the dots on the corresponding graph,  
achieved with contact rates
$\protect\cRbar = \protect\cSbar := b\protect\subS/2 = b\protect\subR/2 = 5$ 
and
$\protect\cIbar := b\protect\subI/2 = 3$ respectively.
}
\end{figure}

For each status $z\in\{\text{S,I,R}\}$ and contact rate $c$ we use
\begin{equation}
u_z^{} (c)  \; =  \; \left( b_z c \, - \, c^2 \right)^g \, - \, a_z,    
\end{equation}
with parameter values 
previously used in \cite{cho2020}:
$g=0.25, \; b\subS = b\subR = 10, \; b\subI = 6; \; a\subS = a\subR = 0, $ and $a\subI = 4.$ 
This specifies the same instantaneous utility for all healthy individuals (status $S$ or $R$) and a different,
generally lower, instantaneous utility for Infected individuals; see Figure~\ref{fig:utilities}.

These utility functions are concave (as long as the shape parameter $g$ is in $(0,1]$) 
and have a unique maximum in contacts, achieved at $c=b_z/2.$ 
Parameter $a_z$ determines the baseline level of utility gain for individuals in state $z$. We naturally enforce $b\subS = b\subR > b\subI$ and $a\subS = a\subR< a\subI$ since infected lose economic productivity and suffer an overall health cost. 
The maximal instantaneous utility values $\uSbar=\uRbar$ and $\uIbar$ are
achieved with contact rates
$\cRbar = \cSbar := b\subS/2 = b\subR/2 = 5$ and $\cIbar := b\subI/2 = 3$ respectively.
These are also the contact rates chosen in the previous section to produce Figure~\ref{fig:noopt200}.

\subsection{Single individual optimizing their Contact Rate Strategy}
\label{ss:optimal_control}
An individual in the population accumulates reward according to the utility function of their current state, 
which is subject to random transitions. 
To begin, suppose that the entire population uses some specific and known contact rate strategy $\bc(t) = (c\subS(t), c\subI(t), c\subR(t)),$ leading to the epidemic trajectory $\y(t) = (S(t), I(t), R(t), D(t))$. If a lone focal individual chooses to follow some different strategy $\bchat = (\hat{c}\subS(t), \hat{c}\subI(t), \hat{c}\subR(t))$, they will not influence the course of the epidemic (i.e., $\y(t)$ will remain the same).
But they will change their personal chances of getting ill and their personal lifestyle. 
Mathematically, for a focal individual starting from a 
state $\xi \in \{S, \, I, \, R, \, D \}$ at the time $t<T$
and using contact rate $\bchat(\cdot)$ until the time $T,$ 
their expected cumulative reward will be
$$
\J \left(\xi, t, \bchat(\cdot); \, \y(\cdot), \bc(\cdot) \right) \; = \; 
\E \left[
\int_t^{T} u_{z(\nt)} \left( \hat{c}_{z(\nt)}(\nt) \right) \, d\nt  
\; + \; \Psi_{z(T)}
\right],
$$
where 
$z(\nt) \in \{S,I,R,D\}$ is that individual's
state at the time $\nt \in [t,T]$,
starting from $z(t) = \xi$ and changing through a continuous in time Markov chain with transition rates based on the current 
state of epidemics and the chosen contact rates.
In addition to their accumulated utility, an individual also receives a terminal reward or penalty $\Psi$ determined by their health status $z(T)$ at the end of this planning horizon.  
Here, $D$ is the absorbing state, with $u\subD \equiv 0$ (no utility accumulated after death) and a large negative terminal ``reward'' $\Psi\subD$
in case of this sad outcome.  If they survive,
it is a natural convention to use $\Psi\subS = \Psi\subR = 0,$ with a non-positive $\Psi\subI$ reflecting possible additional burden for those still infected at the end of the process.
In this paper, we follow \cite{cho2020} and use $\Psi\subI = 0,$ 
which corresponds to assuming that an effective cure for the disease will become available to everyone at the time $T$.
Alternatively, if it is assumed that an effective vaccine will become widely available at the time $T$, a negative $\Psi\subI$ can be used to encode the average 
burden of reduced utility (and the risk of death) experienced after the time $T$ by Infected, 
since they cannot be vaccinated right away with all others \cite{reluga2010game, doebeli2026partial}.

Some of the transition rates governing $z(\nt)$ are constant;
e.g., for the recovery from infection or the loss of immunity to happen within a short time interval $\tau,$
$$
\Pr \left( 
z(\nt + \tau) = R \, \mid \, z(\nt) = I
\right) 
\; = \; \mu\tau \, + \, o(\tau)
\qquad \text{and} \qquad
\Pr \left( 
z(\nt + \tau) = S \, \mid \, z(\nt) = R
\right) 
\; = \; \gamma\tau \, + \, o(\tau).
$$
Other rates are $\y(\nt)$-dependent or even $\left(\y(\nt), \bc(\nt), \bchat(\nt)\right)$-dependent; e.g, for a death to occur within a short time interval $\tau,$
$ \;
\Pr \left( 
z(\nt + \tau) = D \, \mid \, z(\nt) = I
\right) 
\; = \; \delta\!\left(I(\nt)\right)\tau \, + \, o(\tau)$
and for a new infection,
$$
\Pr \left( 
z(\nt + \tau) = I \, \mid \, z(\nt) = S
\right) 
\; = \; \hat{c}\subS (\nt) \, p\!\left( \y(\nt), \bc(\nt) \right) \tau \, + \, o(\tau),
$$
with $p$ determined by the current state of the epidemic and the current contact rates used by all others, as defined in \eqref{eq:original_p}.

Assuming that the focal individual strives to maximize their expected cumulative reward, one can define {\em value functions} to 
quantify the result of optimizing the contact rates:
\begin{equation}
\label{eq:val_def}
V_{\xi} (t) \; = \; \sup\limits_{\bchat(\cdot)}
\J \left(\xi, t, \bchat(\cdot); \, \y(\cdot), \bc(\cdot) \right).
\end{equation}
Here, $\xi \in \{S,I,R\}$ is again their initial state at the time $t \leq T$
and $V_{\xi} (t)$ is the (optimized) expected cumulative reward in the time remaining 
until $T$.
When $z(\nt) = I$ or $z(\nt) = R,$ the focal individual's choices do not influence their transition rates, and thus they will always choose the contact rate that maximizes their current utility function: $\hat{c}\subI^* = \cIbar$
and $\hat{c}\subR^* = \cRbar,$ achieving utility values $\uIbar = \uI(\cIbar)$ and
$\uRbar = \uR(\cRbar)$ respectively; see Figure~\ref{fig:utilities}.
On the other hand, that individual has a non-trivial choice to make when $z(\nt) = S:$
a contact rate lower than $\cSbar$ is  
often optimal since it reduces the immediate chances of getting infected.
Using the above expressions for transition rates, the optimality equations satisfied by the value functions can be written as

\begin{align}
\label{eq:bellman_S}
\VS(t) & \; = \; 
\max_{\hatcS(\cdot)} \left\{\left(\int_t^{t+\tau}
\uS\!\left(\hatcS(\nt) \right) \, d\nt\right) 
\, + \, p \hatcS(t) \tau\VI(t+\tau) 
\, + \, \left(1-p \hatcS(t) \tau \right) \VS(t+\tau)\right\}
\, + \, o(\tau);\\
\label{eq:bellman_I}
\VI(t) & \; = \; \int_t^{t+\tau}\uIbar d\nt 
\, + \, \mu\tau\VR(t+\tau) 
\, + \, \delta\!\left(I(t)\right) \tau\Psi\subD 
\, +\, \left( 1-\mu\tau- \delta\!\left(I(t)\right) \tau \right)\VI(t+\tau)
\, + \, o(\tau);\\
\label{eq:bellman_R}
\VR(t) & \; = \; \int_t^{t+\tau}\uRbar d\nt 
\, + \, \gamma\tau\VS(t+\tau) 
\, + \, (1-\gamma\tau)\VR(t+\tau)
\, + \, o(\tau);
\end{align}
for any starting time $t < T$ and any $\tau \leq T-t.$
Here, $p=p\!\left( \y(t), \bc(t) \right)$ is again computed by formula \eqref{eq:original_p} based on the current state of epidemics.
A standard control-theoretic argument based on a Taylor-series expansion of (\ref{eq:bellman_S}-\ref{eq:bellman_R})
yields a system of ODEs for the value functions. 
For example, focusing on \eqref{eq:bellman_S},
and assuming that $\uS, \cS, \VS$ are smooth,
$$
\int_t^{t+\tau}
\uS\!\left(\hatcS(\nt)\right) \, d\nt 
\; = \; \tau \uS\!\left(\hatcS(t)\right) + o(\tau),
\qquad
p \hatcS(t) \tau\VI(t+\tau)
\; = \; p \hatcS(t) \tau\VI(t) + o(\tau),
$$
$$
\text{and }\qquad
\left(1-p \hatcS(t) \tau \right) \VS(t+\tau)
\; = \; \VS(t) + \tau \VS'(t) - p \hatcS(t) \tau \VS(t) + o(\tau).
$$
For small $\tau,$ this allows us to concentrate on optimizing the {\em current} contact rate $\hatcS(t)$ only.  
Plugging these expression into \eqref{eq:bellman_S}, subtracting $\VS(t)$ from both sides, dividing through by $\tau,$ and taking the limit with $\tau \rightarrow 0,$ we obtain the ODE \eqref{eq:val_ODEs_original_VS} below.  Similar Taylor expansions used in \eqref{eq:bellman_I} and \eqref{eq:bellman_R}
yield ODEs \eqref{eq:val_ODEs_original_VI} and \eqref{eq:val_ODEs_original_VR} respectively.
\begin{align}
\label{eq:val_ODEs_original_VS}
\VS'(t) & \; = \; -\uS(\hatcSS) \, + \, p\!\left( \y(t), \bc(t) \right)
\hatcSS(\VS(t)-\VI(t));\\
\label{eq:val_ODEs_original_VI}
\VI'(t) & \; = \; -\uIbar \, + \, \mu\big(\VI(t) - \VR(t) \big) 
\, + \, \delta \left( I(t) \right) \big(\VI(t) - \Psi\subD\big);\\
\label{eq:val_ODEs_original_VR}
\VR'(t) & \; = \; -\uRbar \, + \, \gamma\big(\VR(t) - \VS(t)\big).
\end{align}
In the above system, we use $\hatcSS$ to denote 
the optimal contact rate while susceptible,
which is  $\left(\y(t), \bc(t) \right)$-dependent and defined by 
\begin{equation}
\label{eq:hatcSS}
\hatcSS \; = \; \argmax_{\hatcS}\left\{
\uS(\hatcS) \, +\,  p\!\left( \y(t), \bc(t) \right)  \hatcS \big( \VI(t) - \VS(t) \big)
\right\}.
\end{equation}
We note that ODEs (\ref{eq:val_ODEs_original_VS}-\ref{eq:val_ODEs_original_VR}) are posed over the time interval $[0,T]$ with the terminal condition $\VS(T)=\VI(T)=\VR(T)=0.$

\subsection{Nash Equilibrium: Mean Field Game System}
\label{ss:Nash}
Of course, if all individuals selfishly and rationally choose their contact rate, this will alter the course of the epidemic, with optimality thus replaced by the notion of Nash equilibrium.
(Given the constant transition rates from $I$ and $R,$ it is easy to see that 
all rational selfish agents will still use the utility-maximizing contact rates $\cIbar$ and $\cRbar$ while Infected and Recovered respectively.  But the choice of optimal contact rate while Susceptible will now become more subtle since it will affect both $\y(\cdot)$ and their immediate chances of getting Infected.) 
As usual in MFG literature,  
we will say that 
$\bc(t) = \left(c\subS(t), c\subI(t), c\subR(t) \right)$ 
and
$\y(t) = \left(S(t), I(t), R(t), D(t) \right)$
form a Nash equilibrium if\\[1em]
(1) $\quad \y(t)$ results from $\bc(t)$ via ODEs 
(\ref{eq:SIRSD_ODEs_original_S}-(\ref{eq:SIRSD_ODEs_original_D}) and\\[1em]
(2) 
$\quad \J \left(\xi, t, \bc(\cdot); \, \y(\cdot), \bc(\cdot) \right) \; \geq \;
\J \left(\xi, t, \bchat(\cdot); \, \y(\cdot), \bc(\cdot) \right)$\\
for all 
$\xi \in \{S, \, I, \, R, \, D \}, \; t\in[0,T], \;$
and all contact rate policies $\bchat(\cdot).$
The latter condition ensures that no individual has any incentive for switching their personal contact rate.

Using $p(\cS)$ as a shorthand for
\begin{equation}
\label{eq:MFG_p}
p \!\left(S, I, R; \cS \right) \; = \;
\beta 
\frac{\cIbar I}{\cS S \, + \, \cIbar I \, + \, \cRbar R},
\end{equation}
the resulting Mean Field Game can be now described by a system of seven ODEs: 
\begin{align}
\label{eq:MFG_ODEs_original_S}
S' &\; = \; - p\!\left(\cSS\right) \cSS S \, + \, \gamma R;\\
\label{eq:MFG_ODEs_original_I}
I' &\; = \;  p\!\left(\cSS\right) \cSS S \, - \, \mu I \, - \, \delta(I) I;\\
\label{eq:MFG_ODEs_original_R}
R' &\; = \;  \mu I \, - \, \gamma R;\\
\label{eq:MFG_ODEs_original_D}
D' &\; = \;  \delta(I) I;\\
\label{eq:MFG_ODEs_original_VS}
\VS' & \; = \; -\uS(\cSS) \, + \, p\!\left( \cSS \right)
\cSS(\VS-\VI);\\
\label{eq:MFG_ODEs_original_VI}
\VI' & \; = \; -\uIbar \, + \, \mu\big(\VI - \VR \big) 
\, + \, \delta(I) \big(\VI - \Psi\subD\big);\\
\label{eq:MFG_ODEs_original_VR}
\VR' & \; = \; -\uRbar \, + \, \gamma\big(\VR - \VS\big);
\end{align}
with the Nash contact rate of susceptible individuals specified by
\begin{equation}
\label{eq:cSS}
\cSS \; = \; \argmax_{\cS}\left\{
\uS(\cS) \, +\,  p\!\left( c\subS^{\newest{*}} \right)  \cS \big( \VI - \VS \big)
\right\}.
\end{equation}

\begin{remark}
To aide the readers less familiar with MFG-type models, we emphasize the distinction,
which is obvious for MFG experts.  Despite the many similarities, the above model
is very different from the focal-individual discussion in the previous subsection.
In \S \ref{ss:optimal_control}, we assumed that the ODEs (\ref{eq:SIRSD_ODEs_original_S}-(\ref{eq:SIRSD_ODEs_original_D})
were first solved forward in time using the known contact rates $\bc(t)$
adopted by the entire population.  The focal individual was then trying to find 
a better contact rate policy $\bchat(t)$ (that would be used by them and them alone)
by solving the ODEs (\ref{eq:val_ODEs_original_VS}-\ref{eq:val_ODEs_original_VR})
backward in time
with the epidemic trajectory $\y(t) = \left(S(t), I(t), R(t), D(t) \right)$ already fixed
based on $\bc(t).$  In contrast, the above ODE system 
(\ref{eq:MFG_ODEs_original_S}-\ref{eq:MFG_ODEs_original_VR})
is fully coupled via 
\eqref{eq:cSS} also influencing $\y(t)$, and thus should be treated as a Two Point Boundary Value Problem (TPBVP) with initial conditions specified for $S,I,R,D$ and terminal conditions specified for $\VS,\VI,\VR.$

\newest{
Another subtlety worth noting here is that our definition of Susceptibles' optimal contact rate is implicit since $\cSS$ is also used 
on the right hand side of formula \eqref{eq:cSS}.  This is consistent with our assumption that all agents are ``small'' and independent.  Thus, when searching for their best personal contact rate, they do not cooperate, and each of them does not expect others to necessarily follow their lead.  As a result, when optimizing their choices,  each views the risk of getting infected per contact as $p(\cSS)$ (based on the group behavior) rather than $p(\cS)$ (based on their personal preferred contact rate).  The strict concavity of our utility function $\uS$ ensures that the argmax in
$$
F(c) := \argmax_{\cS}\left\{
\uS(\cS) \, +\,  p\!\left( c \right)  \cS \big( \VI - \VS \big)
\right\}
$$
is always a singleton for every population-wide susceptible contact rate
$c \in [0, \cSbar],$ and our $\cSS$ is the fixed point of this best-response operator $F$.  This nuance is entirely due to our frequency-based force of infection and
does not arise in MFG-epidemiological models that assume a more common density-based force of infection; see, e.g., \cite{reluga2010game, cho2020, doebeli2026partial}. 
}

In \newest{general} MFG literature, such systems \newest{with initial and terminal conditions}
are often treated numerically by
forward-backward iterations, which, for stability reasons, might also be combined with 
an under relaxation \cite{lauriere2021numerical} or with the so called ``fictitious play'' \cite{cardaliaguet2017learning}.
Since our models in this paper are over a finite state space, we instead employ the standard collocation methods for TPBVPs (e.g., the built-in Matlab function {\tt bvp5c} \cite{kierzenka2008bvp}).
The initial guess for the TPBVP solver is generated by 
solving the ``single optimizer problem'' 
(\ref{eq:val_ODEs_original_VS}-\ref{eq:val_ODEs_original_VR})
for the epidemics trajectory corresponding
to a population-wide use of the specific
contact rate policy $\bc(t) = (\cSbar, \cIbar, \cRbar).$
\end{remark}

\subsection{B-MFG: adding behavioral patterns and switching}
\label{ss:multi_pop_MFG}
The model described above is close in spirit to others commonly presented in MFG literature; e.g., see \cite{cho2020} and \cite{doebeli2026partial} for similar models but with the density-based force of infection.
We now extend it to reflect three different behavioral patterns, modeled as three distinct subpopulations: Reckless/stubborn, Followers/consensus-seekers, and Optimizers.
All of them use the same contact rates while Infected (i.e., $\cIbar$) or Recovered/Immune (i.e., $\cRbar$) since these rates maximize their instantaneous utility function and do not influence the rate of transition to other health states.
Thus, subpopulations are only different in how they choose their contact rates while Susceptible, but we maintain three health compartments ($S$, $I$, and $R$) in each subpopulation separately, using the subscripts ($r$, $f$, and $o$) to distinguish them\footnote{This is necessary because we do not assume that one's behavioral pattern/subpopulation changes automatically due to a change in one's health status:
a reckless individual may not become more careful just because they suffered from the infection.  In fact, some of the studies show that those recovered from a disease
are in some cases more risk-prone afterwards \cite{egeli2021risk, kaim2022we,  huang2023covid}.}.
The same subscript notation is also used to distinguish the contact rates and the value functions.  
$S$, $I$, $R$ variables without subscript will now refer to the total number of people in each respective class across all sub-populations; e.g., $S = S_r + S_f + S_o.$ 

Susceptible Reckless individuals ignore the epidemiological situation and the risk of getting infected, thus selecting the contact rate that myopically maximizes their utility function: $\; c\subSr(t) \, = \, \cSbar.$ 

Susceptible Followers adopt the ``consensus'' modeled by averaging the contact rate used by all Susceptibles (across all subpopulations).
This averaging will obviously depend on the specific contact rate $\cSoS(t)$ selected by the Susceptible Optimizers through the process described below.  But for now we simply characterize the Follower's chosen rate as a function of whatever contact rate $c$ those Optimizers might be considering:    
\begin{equation}
\label{eq:c_subSf}
c\subSf(c) 
\; = \; 
\frac{\cSbar S_r \, + \, c\subSf(c) S_f  \, + \, c S_o}{S}    
\; = \; \frac{ \cSbar S_r \, + \, c S_o}{S_r \, + \, S_o},    
\end{equation}
where  
the second (explicit) expression for $c\subSf(c)$ is valid as long as $S_r(t) \, + \, S_o(t) \neq 0$ (otherwise the Followers don't have anyone to follow). 

Finally, the rational Susceptible Optimizers will select a Nash-type contact rate 
$c\subSo^*$, but taking into account two new complications:
\begin{enumerate}
    \item 
the behavior of other subpopulations (including the fact that the Followers will be immediately influenced by whatever the Optimizers select) and 
\item 
the possibility that all people (including the Optimizers!) might occasionally change their behavioral patterns in the future. (E.g., an Optimizer becoming Reckless for a while or vice versa.)
\end{enumerate}
We note that the latter reflects a subtle shift in modeling assumptions.  
In MFG literature, optimizing agents are assumed to be not only fully rational in formulating their Nash-equilibrium strategies, but also fully {\em consistent} in implementing those strategies long term.
Instead, we assume that Optimizers are inconsistent (prone to possible/occasional behavioral switches) and understand that they will follow the rationally chosen strategy only while in that Optimizer mode.  
They are further assumed to use this self-awareness when searching for an optimal (Nash-equilibrium) strategy.

To address the first of these complications,
we need to modify our formula for computing the per contact probability of 
becoming infected. Extending  
\eqref{eq:MFG_p}, we will now use
$p(c)$ as a shorthand for 
\begin{equation}
\label{eq:MFG_p_extended}
p \!\left(S_r, S_f, S_o, I, R; c \right) \; = \;
\beta 
\frac{\cIbar I}{\cSbar S_r \, + \, c\subSf\left(c\right) S_f \, + \, c S_o \, + \, \cIbar I \, + \, \cRbar R},
\end{equation}
where $c$ is the contact rate currently considered by Susceptible Optimizers.

As for the behavioral switches, we assume that they are 
happening at random times, driven by a nonhomogeneous Poisson process with rates $\lambda_{j,k},$ which are possibly dependent
on the changing epidemiological situation and  contact rates.
More precisely, we assume that for an individual who is currently
using a behavioral pattern $j \in \left\{r, f, o \right\}$
the probability of switching to another behavioral pattern 
$k \in \left\{r, f, o \right\} \backslash \left\{j\right\}$ 
within a short time $\tau$ is equal to $\lambda_{j,k} \tau + o(\tau).$
The fact that $\lambda_{j,k}$ are not necessarily constant is a natural modeling assumption; e.g., transitions to reckless behavior are less likely when the total number of Infected people is high and more likely when Nash-style rational behavior leads to fatigue or life-style envy (when  $u\subS\left(c\subSo^*\right)$ is much lower than $\uSbar$).  
For the sake of notational simplicity, we mostly omit these dependencies of  $\lambda_{j,k}$, though we do keep them in mind in stating the optimization problem. They are also illustrated in some of the computational experiments in \S \ref{s:results}.

The value function for a Susceptible Optimizer, representing the expected rewards for the remaining time interval $[t,T],$ will satisfy the following optimality equation for any sufficiently small $\tau$:

\begin{align}
\nonumber
V\subSo(t) & \; = \; 
\max_{\cS(\cdot)} \left\{\left(\int_t^{t+\tau}
\uS\!\left(\cS(\nt) \right) \, d\nt\right) 
\, + \, p \cS(t) \tau V\subIo(t+\tau) 
\, + \, \lambda_{o,r} \tau V\subSr(t+\tau)
\, + \, \lambda_{o,f} \tau V\subSf(t+\tau)
\right.\\
& \hspace*{52mm} \left. \, + \,  
\left(1 \, - \, 
\left( p \cS(t) + \lambda_{o,r} + \lambda_{o,f} \right) \tau \right) V\subSo(t+\tau)
\right\}
\, + \, o(\tau),
\end{align}
where the dependence of $p, \lambda_{o,r},$ and $\lambda_{o,f}$
on $\cS(t)$ are not indicated explicitly to keep the notation manageable.
This equation is a generalization of \eqref{eq:bellman_S}, and a similar Taylor series expansion argument yields an ODE for the value function
\begin{equation}
\label{eq:MFG_ODEs_extended_VSo}
V\subSo' \; = \; -\uS(\cSoS) \, +\, p(\cSoS) \cSoS \big( V\subSo - V\subIo \big)
\,+\, \lambda_{o,r}(\cSoS) \big( V\subSo - V\subSr \big)
\,+\, \lambda_{o,f}(\cSoS) \big( V\subSo - V\subSf \big)
\end{equation}
with the optimal contact rate found from
\begin{equation}
\label{eq:cSoS}
\cSoS \; = \; \argmax_{c} \left\{
\uS(c) \, +\, c \, p(\newest{\cSoS})  \big( V\subIo  - V\subSo  \big)
\,+\, \lambda_{o,r}(c) \big( V\subSr - V\subSo \big)
\,+\, \lambda_{o,f}(c) \big( V\subSf - V\subSo \big)
\right\}.
\end{equation}

The resulting B-MFG model can be now formally described by a system of 19 coupled ODEs
using indices 
$j \in \left\{r, f, o \right\}$
and  
$k \in \left\{r, f, o \right\} \backslash \left\{j\right\}$ 
: 
\begin{align}
\label{eq:MFG_ODEs_extended_S}
S_j' &\; = \; - p\!\left(\cSoS\right) \cSj S_j \, + \, \gamma R_j
 \, + \, \sum\limits_{k \neq j}
 \left(
 \lambda_{k,j} S_k -  \lambda_{j,k} S_j 
 \right);\\
\label{eq:MFG_ODEs_extended_I}
I_j' &\; = \;  p\!\left(\cSoS\right) \cSj S_j \, - \, \mu I_j \, - \, \delta(I) I_j
 \, + \, \sum\limits_{k \neq j}
 \left(
 \lambda_{k,j} I_k -  \lambda_{j,k} I_j 
 \right);\\
\label{eq:MFG_ODEs_extended_R}
R_j' &\; = \;  \mu I_j \, - \, \gamma R_j
 \, + \, \sum\limits_{k \neq j}
 \left(
 \lambda_{k,j} R_k -  \lambda_{j,k} R_j 
 \right);\\
\label{eq:MFG_ODEs_extended_D}
D' &\; = \;  \delta(I) I;\\
\label{eq:MFG_ODEs_extended_VS}
V\subSj' & \; = \; -\uS(\cSj) \, + \, p\!\left( \cSoS \right)
\cSj(V\subSj-V\subIj)
 \, + \, \sum\limits_{k \neq j}
 \lambda_{j,k}
 \left(
 V\subSj - V\subSk
 \right);\\
\label{eq:MFG_ODEs_extended_VI}
V\subIj' & \; = \; -\uIbar \, + \, \mu\big(V\subIj - V\subRj \big) 
\, + \, \delta(I) \big(V\subIj - \Psi\subD\big)
 \, + \, \sum\limits_{k \neq j}
 \lambda_{j,k}
 \left(
 V\subIj - V\subIk
 \right);\\
\label{eq:MFG_ODEs_extended_VR}
V\subRj' & \; = \; -\uRbar \, + \, \gamma \big(V\subRj - V\subSj\big)
 \, + \, \sum\limits_{k \neq j}
 \lambda_{j,k}
 \left(
 V\subRj - V\subRk
 \right);
\end{align}
where
$c\subSr^{} = \cSbar, \, c\subSo^{} = \cSoS$
is specified by formula \eqref{eq:cSoS}, and 
$c\subSf^{} = c\subSf^{}\left(\cSoS\right)$ is specified by 
formula \eqref{eq:c_subSf}.
For the sake of readability,
we do not explicitly list above the dependency
of behavioral switching rates $\lambda$ on $\cSoS$ and $I$.
\newest{But if they are not constants, both $\lambda_{o,r}$ and $\lambda_{o,f}$
are assumed to be non-increasing in $I$ and convex in $\cSoS.$ 
}
This is again a TPBVP, with initial conditions specified for
$\left(S_j, I_j, R_j\right)_{j \in \{r,f,o\}}$ and $D$, while zero terminal conditions are assumed
for $\left(V\subSj, V\subIj, V\subRj \right)_{j \in \{r,f,o\}}.$

\section{Results}
\label{s:results}
\begin{figure}[ht]
\centering

\begin{subfigure}{\textwidth}
  \centering
  \begin{minipage}{0.8cm}
    \refstepcounter{subfigure}\subcaption*{(\thesubfigure)} 
  \end{minipage}
  \begin{minipage}{12cm}
    \includegraphics[height=4cm]{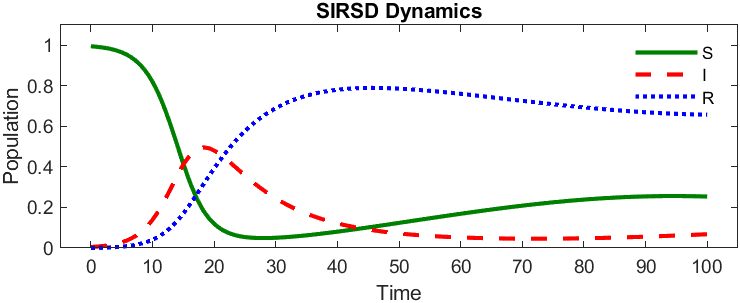}
  \end{minipage}%
  
\end{subfigure}

\begin{subfigure}{\textwidth}
  \centering
  \begin{minipage}{0.8cm}
    \refstepcounter{subfigure}\subcaption*{(\thesubfigure)} 
  \end{minipage}
  \begin{minipage}{12cm}
  
  \vspace{0.07cm}
    \includegraphics[height=4cm]{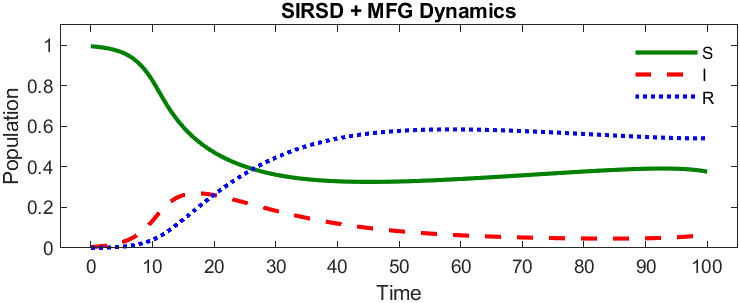}
  \end{minipage}%
\end{subfigure}
\hspace*{0.12cm}
\begin{subfigure}{\textwidth}
  \centering
  \begin{minipage}{0.8cm}
    \refstepcounter{subfigure}\subcaption*{(\thesubfigure)} 
  \end{minipage}
  \begin{minipage}{12cm}
 \includegraphics[height=4cm]{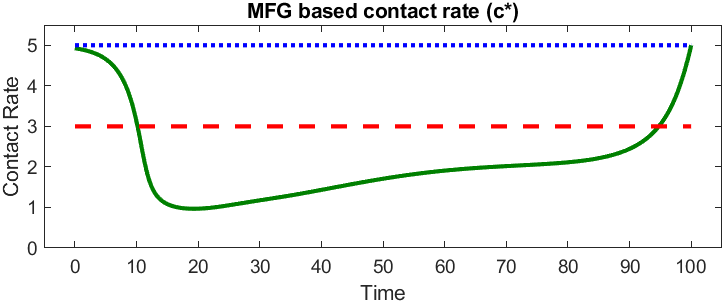}
   
  \end{minipage}%
  
\end{subfigure}

\caption{ Basic SIRSD epidemic dynamics (top) and mean field game epidemic dynamics (middle) for an initial infected population of $I_0$ = 0.005 or 0.5\%. The optimal contact rate for the optimizing populations in the MFG + SIRSD model is displayed in the bottom row. Computed for the time horizon $T=100.$}
\label{fig:sirsd}
\end{figure}

We note that the model summarized by equations (\ref{eq:MFG_ODEs_extended_S}-\ref{eq:MFG_ODEs_extended_VR}) is fairly broad.  In particular, with the fully Reckless population and no behavioral switches 
(i.e., $S_k(0) = I_k(0) = R_k(0) = 0$ and $\lambda_{r,k} = 0$ for all $k \in \{f, \, o\}$)
it reduces to the traditional SIRSD model described in \S\ref{ss:baseline_model} and illustrated in Figure \ref{fig:noopt200}.
We begin by comparing it with the other extreme case:
the ``traditional MFG'' scenario --
a fully rational population of Optimizers and no behavioral switches
(i.e., $S_k(0) = I_k(0) = R_k(0) = 0$ and $\lambda_{o,k} = 0$ for all $k \in \{r, \, f\}$), illustrated 
in Figure \ref{fig:sirsd}.  In both cases, we use the time horizon $T=100$
and start with $0.5\%$ of population Infected and all others initially Susceptible.
The middle graph shows that Optimizers, acting independently and optimizing their contact rates selfishly, still manage to flatten the curve of the epidemic and to reduce the peak  
\newest{percentage} 
of Infected 
\newest{substantially (from 49.6\% to 27\%).} 
While we are not plotting $D(t),$
there is also a reduction in the total \newest{percentage} of deaths $D(T)$: 
from 2.08\% in a fully Reckless population to \newest{1.42}\% in a population consisting of fully rational Optimizers.
This is achieved through a reduction in Susceptible Optimizers' contact rate (bottom row of Figure \ref{fig:sirsd}), which varies with the danger (based on the current number of Infected) and the time remaining until $T$.  We note that the increase in contact rate towards the end (and the corresponding slight increase in Infected) is consistent with the lack of terminal penalty
for Infected (i.e., $V\subIj(T)=0$)
-- a consequence of our assumption that
an effective cure becomes available at the time $T$. (Susceptibles are less afraid of becoming Infected in the last $(1/\mu)$ days since they will not stay ill as long as usual.)

\subsection{Sensitivity to parameters in behaviorally homogeneous populations}
 
Before introducing interacting sub-populations, we compare the sensitivity of the epidemic trajectory to parameter variations in these same two ``homogeneous population'' scenarios: Reckless only (or the original SIRSD model, i.e., 
$S_k(0) = I_k(0) = R_k(0) = 0$ and $\lambda_{r,k} = 0$ for all $k \in \{o, \, f\}$)
and Optimizers only
(i.e., $S_k(0) = I_k(0) = R_k(0) = 0$ and $\lambda_{o,k} = 0$ for all $k \in \{r, \, f\}$).

\begin{figure}[ht]
\centering
\setcounter{subfigure}{0}

\begin{subfigure}{\textwidth}
  \centering
    \begin{minipage}{0.5cm} 
    \refstepcounter{subfigure}%
    \makebox[\linewidth][l]{(\thesubfigure)} 
  \end{minipage}
  \begin{minipage}{12cm}
    \includegraphics[height=4cm]{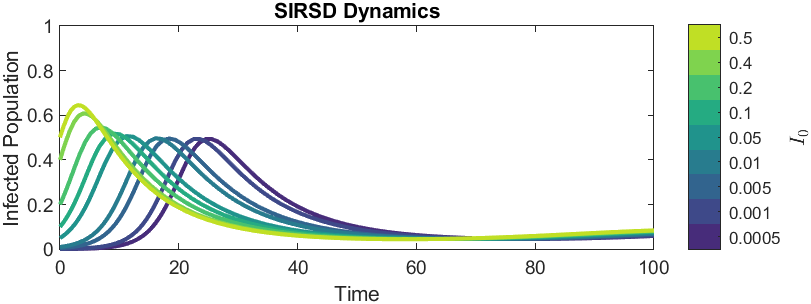}
  \end{minipage}

\end{subfigure}

\begin{subfigure}{\textwidth}
  \centering
  \begin{minipage}{0.5cm}
    \refstepcounter{subfigure}%
    \makebox[\linewidth][l]{(\thesubfigure)}
  \end{minipage}
  \begin{minipage}{12cm}
  
  \vspace{0.07cm}
    \includegraphics[height=4cm]{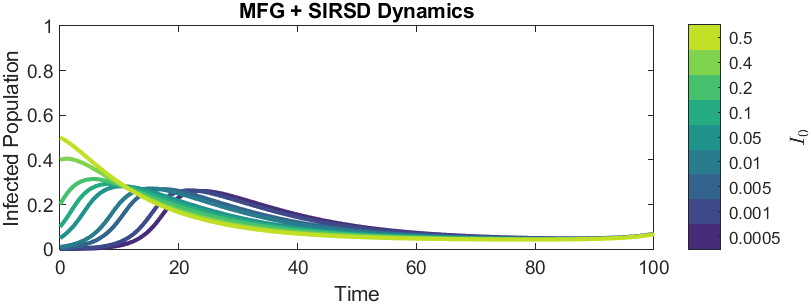}
  \end{minipage}
\end{subfigure}

\begin{subfigure}{\textwidth}
  \centering
  \begin{minipage}{0.5cm}
    \refstepcounter{subfigure}%
    \makebox[\linewidth][l]{(\thesubfigure)}
  \end{minipage}
  \hspace{0.3cm}
  \begin{minipage}{12cm}
    \includegraphics[height=4cm]{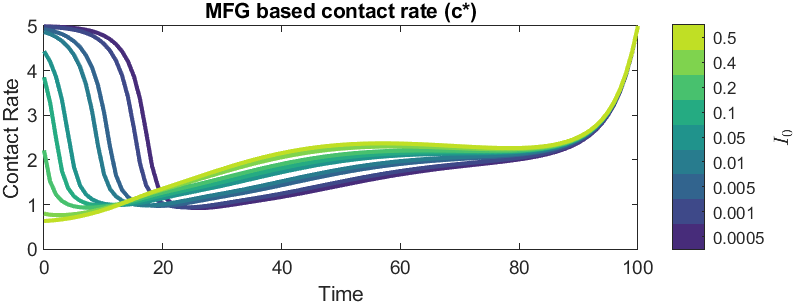}
    \end{minipage}
  
\end{subfigure}

\caption{Varying $I_0$, the initial proportion of infected in the population.
Infected populations for SIRSD (top) and MFG+SIRSD (middle) and the optimal contact rates (bottom) for various $I_0.$}
\label{fig:xi}
\end{figure}
We first vary the initial proportion of infected in the population; see Figure~\ref{fig:xi}.
In both cases, increasing $I_0$ causes higher and earlier peaks in infections.  Comparing the two scenarios, the peaks are generally lower and happen earlier in the case of Optimizers, since their strategic/selfish behavior decreases the effective reproductive number.  Interestingly, for relatively low $I_0,$ the Susceptible Optimizers' contact rate remains high at first (to avoid impacting their lifestyle satisfaction/instantaneous utility while the probability of catching the infection is still fairly low) and decreases rapidly once the proportion of Infected reaches some threshold.

\begin{figure}[ht]
\centering
\setcounter{subfigure}{0}

\begin{subfigure}{\textwidth}
  \centering
  \begin{minipage}{0.5cm} 
    \refstepcounter{subfigure}%
    \makebox[\linewidth][l]{(\thesubfigure)} 
  \end{minipage}
  \begin{minipage}{12cm}
    \includegraphics[height=4cm]{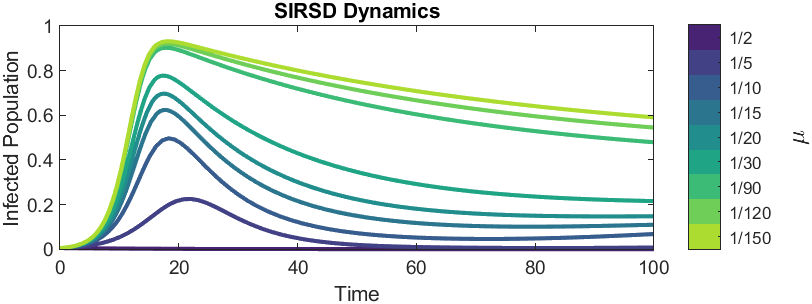}
  \end{minipage}
  
\end{subfigure}

\begin{subfigure}{\textwidth}
  \centering
  \begin{minipage}{0.5cm}
    \refstepcounter{subfigure}%
    \makebox[\linewidth][l]{(\thesubfigure)}
  \end{minipage}
  \begin{minipage}{12cm}
  
  \vspace{0.07cm}
    \includegraphics[height=4cm]{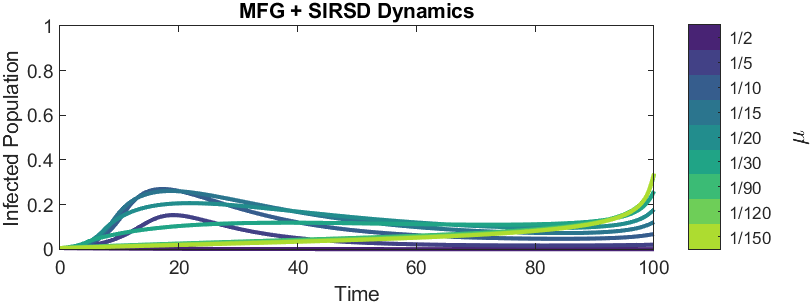}
  \end{minipage}
  
\end{subfigure}

\begin{subfigure}{\textwidth}
  \centering
  \begin{minipage}{0.5cm}
    \refstepcounter{subfigure}%
    \makebox[\linewidth][l]{(\thesubfigure)}
  \end{minipage}
  \hspace{0.3cm}
  \begin{minipage}{12cm}
    \includegraphics[height=4cm]{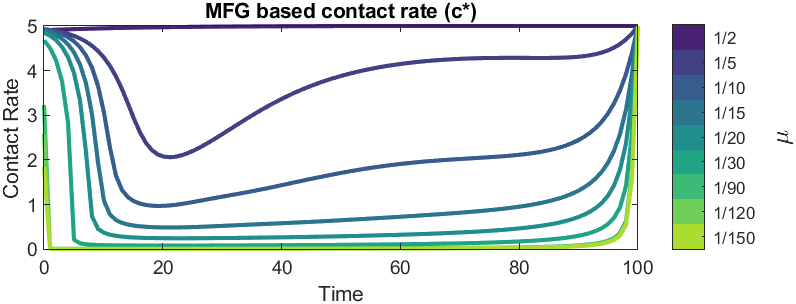}
  \end{minipage}
  
\end{subfigure}

\caption{Varying the recovery rate $\mu$: a comparison of ``Reckless only'' and ``Optimizers only'' scenarios.  
Same format as in Fig.~\ref{fig:xi}.
The presented results correspond to the $\gamma = 1/90$ vertical line in Figure \ref{fig:gamma_mu_heatmaps}.}
\label{fig:mu}
\end{figure}
Turning to dependence on the recovery rate $\mu$ (Figure \ref{fig:mu}), we now see a significant qualitative difference between the two scenarios.  In the traditional SIRSD (or ``Reckless only'') model, the peak of infection increases monotonically as $\mu$ decreases, since Infected individuals now have a longer time to infect others. 
A more complicated and largely the opposite trend holds for the population of Optimizers. 
Ignoring the late upswing just before the cure becomes available, the \newest{initial} peak of infection first increases as $\mu$ varies from $\frac{1}{2}$ to $\frac{1}{10},$ and the reasons for this are the same as in the above ``traditional SIRSD'' scenario. 
But from there on (as $\mu$ decreases further and the expected time to recovery increases correspondingly), the peak of infection decreases significantly  
because Susceptible Optimizers respond to the risk of prolonged illness
with increasingly drastic reductions in contact rates.
This phenomenon is also consistent with prior observations reported in \cite{cho2020}
for an MFG-based SIR model with a density-based force of infection, no immunity waning, and no deaths.

\begin{figure}[ht]
\centering
\setcounter{subfigure}{0}

\begin{subfigure}{\textwidth}
  \centering
  \begin{minipage}{0.5cm}
    \refstepcounter{subfigure}%
    \makebox[\linewidth][l]{(\thesubfigure)} 
  \end{minipage}
  \begin{minipage}{12cm}
    \includegraphics[height=4cm]{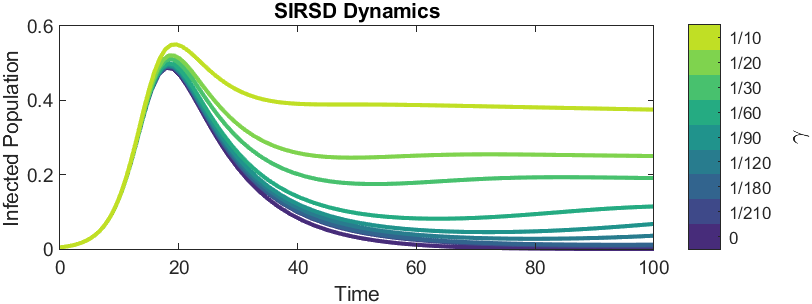}
  \end{minipage}
  
\end{subfigure}

\begin{subfigure}{\textwidth}
  \centering
  \begin{minipage}{0.5cm}
    \refstepcounter{subfigure}%
    \makebox[\linewidth][l]{(\thesubfigure)}
  \end{minipage}
  \begin{minipage}{12cm}
  
  \vspace{0.07cm}
    \includegraphics[height=4cm]{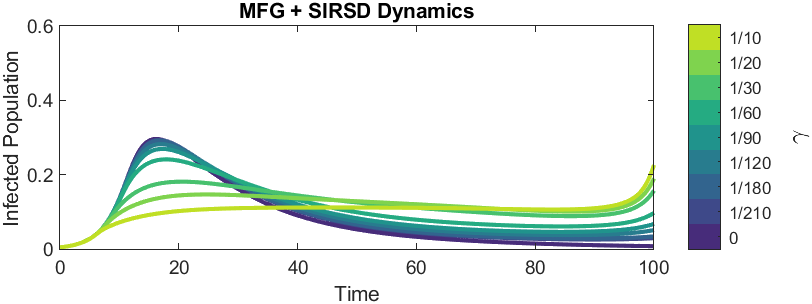}
  \end{minipage}
\end{subfigure}

\begin{subfigure}{\textwidth}
  \centering
  \begin{minipage}{0.5cm}
    \refstepcounter{subfigure}%
    \makebox[\linewidth][l]{(\thesubfigure)}
  \end{minipage}
  \hspace{0.3cm}
  \begin{minipage}{12cm}
    \includegraphics[height=4cm]{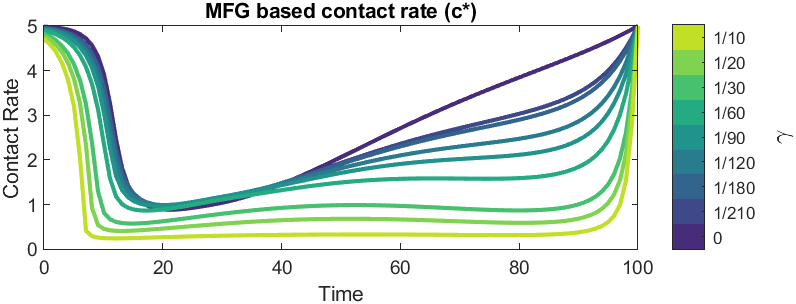}
  \end{minipage}
  
\end{subfigure}
\caption{ Varying the immunity waning rate $\gamma$: a comparison of ``Reckless only'' and ``Optimizers only'' scenarios.  Same format as in Fig.~\ref{fig:xi}.
The presented results correspond to the $\mu = 1/10$ horizontal line in Figure \ref{fig:gamma_mu_heatmaps}.}
\label{fig:gamma}
\end{figure}
A similar contrast is also clear when we vary the rate of immunity waning, $\gamma;$ see Figure~\ref{fig:gamma}.
In traditional SIRSD models, shorter immunity (and larger $\gamma$) leads to more infected individuals: 
a slightly higher peak and, much more noticeably, a higher $I(t)$ long after that peak due to re-infections of many newly susceptibles.
The latter is also true for the population of Optimizers, but the peak clearly decreases since it is driven by   lower contact rates of Susceptible Optimizers, adopted in response to larger $\gamma.$ 

\begin{figure}[ht]
    \arraycolsep=8pt
    \centering
    $
    \begin{array}{cc}
    \includegraphics[height=10cm]{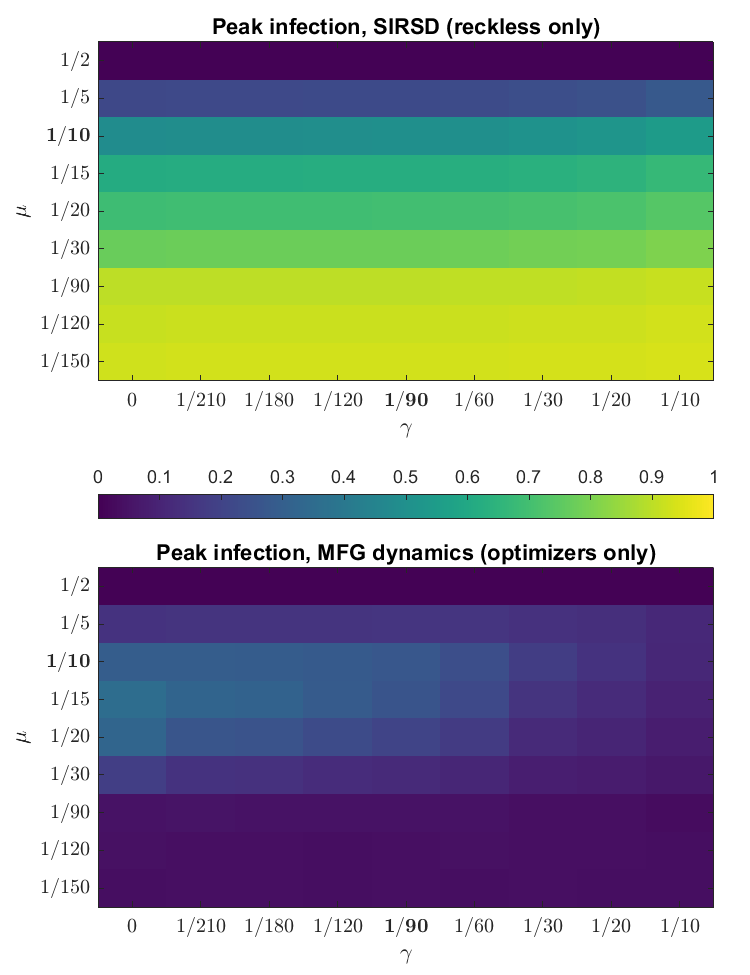}
    &
    \includegraphics[height=10cm]{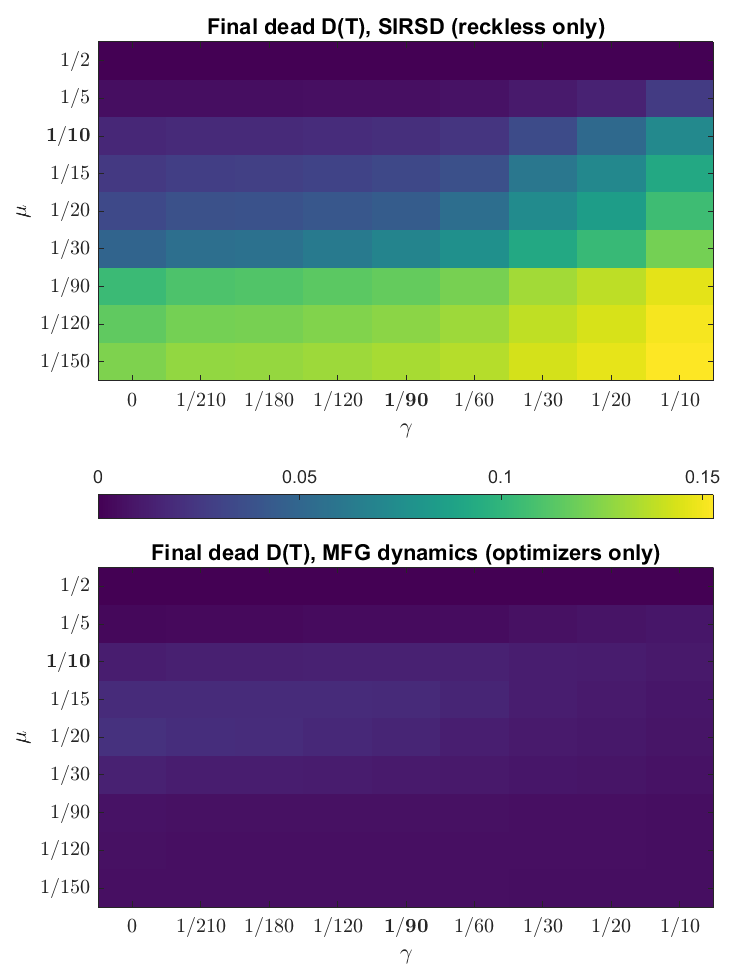}
    \end{array}
    $
\caption{Epidemics outcomes varying in response to changes in parameters
$\gamma$ (the immunity waning rate) and $\mu$ (the recovery rate).
The peak of infection (defined as $\max_{t \in [0, T/2]} I(t)$) is shown on the left
while the total number of dead $D(T)$ is shown on the right.  
The top row corresponds to the ``classical SIRSD'' dynamics, 
which is equivalent to the ``everyone is Reckless'' scenario in our model.
The bottom row corresponds to the ``pure MFG-SIRSD'' dynamics, which is equivalent 
to the ``everyone is an Optimizer'' scenario in our model.
The axis are not uniformly scaled, with the sampled parameter values corresponding to epidemics trajectories shown in subsequent figures.  In particular, \newest{the column $\gamma = 1/90$ corresponds to Fig.~\ref{fig:mu} while the row $\mu = 1/10$ corresponds to Fig.~\ref{fig:gamma}.}
}
\label{fig:gamma_mu_heatmaps}
\end{figure}

To provide a more intuitive summary of $(\mu, \gamma)$-dependence, 
we also include heatmaps in Figure~\ref{fig:gamma_mu_heatmaps},
which show how varying these parameters affects the two main 
epidemics outcomes: the peak infection fraction\footnote{
As Figures~\ref{fig:mu} and \ref{fig:gamma} show,
since everyone knows that the cure will be available at the time $T$, extremely low values of $\mu$ and extremely high values of $\gamma$ can lead to late infection outbreaks under MFG dynamics. To ensure a fair comparison across a broad range of $(\mu, \gamma)$
values, in Figure~\ref{fig:gamma_mu_heatmaps}
we define the ``peak infection fraction'' as $\max_{t \in [0, T/2]} I(t),$
computing this performance measure for the initial wave of the epidemic.}
 and the total number of dead $D(T)$.
Overall, the heatmaps show predictably monotone dependence of outcomes on these parameters for the classical SIRSD model and much better outcomes (but with a clear non-monotonicity) in the MFG model due to  Optimizers' strategic behavior. 

\begin{figure}[ht]
\centering

\vspace*{-5mm}
\begin{subfigure}{\textwidth}
  \centering
  \begin{minipage}{0.5cm}
    \refstepcounter{subfigure}\subcaption*{(\thesubfigure)} 
  \end{minipage}
  \begin{minipage}{12cm}
  
    \includegraphics[height=4cm]{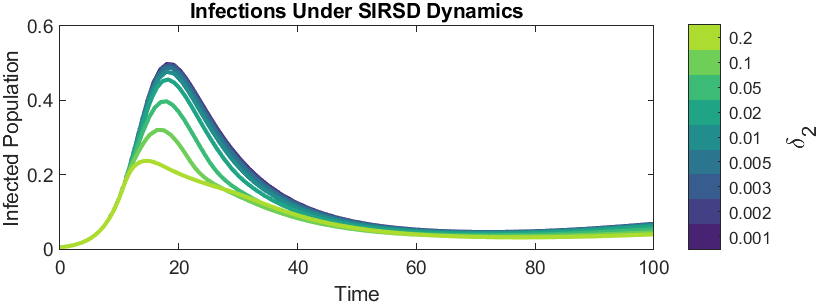}
  \end{minipage}%
  
\end{subfigure}

\begin{subfigure}{\textwidth}
  \centering
 \begin{minipage}{0.5cm}
    \refstepcounter{subfigure}\subcaption*{(\thesubfigure)} 
  \end{minipage}
  \begin{minipage}{12cm}
  \vspace{0.1cm}
    \includegraphics[height=4cm]{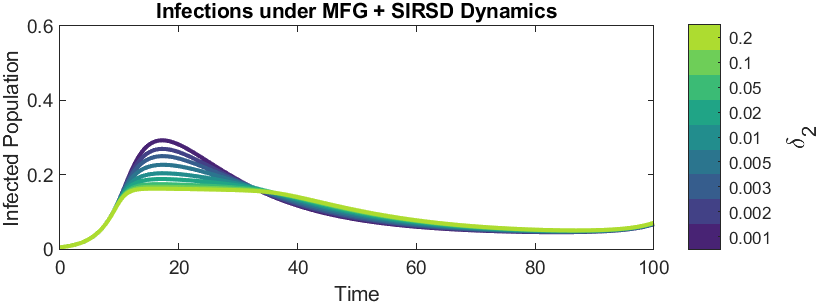}
  \end{minipage}
\end{subfigure}

\begin{subfigure}{\textwidth}
  \centering
  \begin{minipage}{0.5cm}
    \refstepcounter{subfigure}\subcaption*{(\thesubfigure)}
  \end{minipage}
  \hspace{0.3cm}
  \begin{minipage}{12cm}
    \includegraphics[height=4cm]{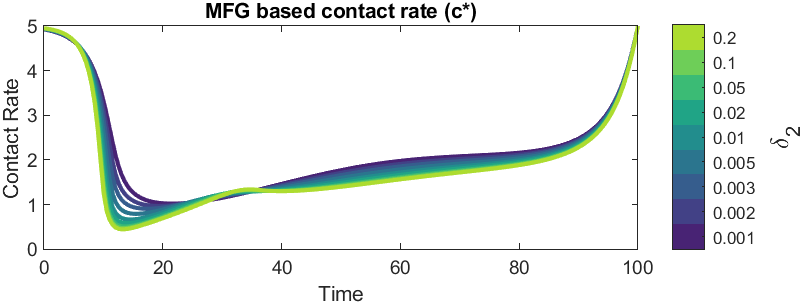}
  \end{minipage}
\end{subfigure}

\begin{subfigure}{\textwidth}
  \centering
  \begin{minipage}{0.5cm}
    \refstepcounter{subfigure}\subcaption*{(\thesubfigure)}
  \end{minipage}
  \hspace{0.1cm}
  \begin{minipage}{12cm}
  
  \vspace{0.07cm}
    \includegraphics[height=4cm]{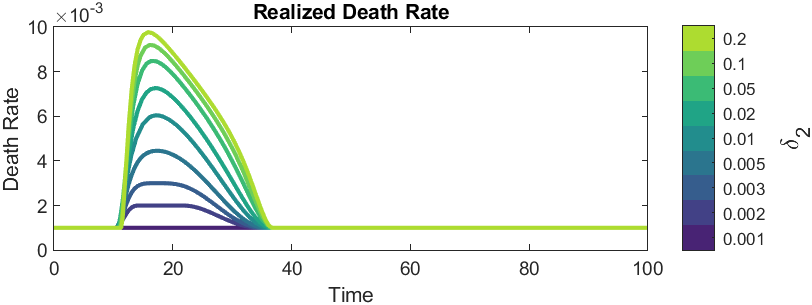}
  \end{minipage}
\end{subfigure}

\begin{subfigure}{\textwidth}
  \centering
  \begin{minipage}{0.5cm}
    \refstepcounter{subfigure}\subcaption*{(\thesubfigure)}
  \end{minipage}
  \begin{minipage}{12cm}
  \vspace{0.1cm}
    \includegraphics[height=4cm]{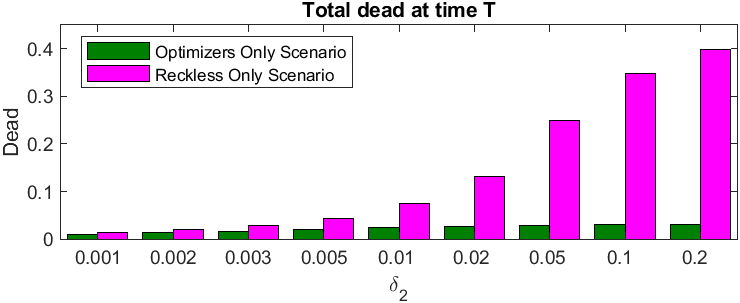}
  \end{minipage}
\end{subfigure}
\caption{Varying the high-mortality rate $\delta_2$, which is only in effect when the infections are high ($I>d_2$). 
(a) Infected as a function of time for the Reckless-only scenario.
The next three subfigures correspond to the Optimizers-only scenario showing
the time-dependent number of Infected in (b), contact rate of Susceptible Optimizers
in (c), and the death rate $\delta(I(t))$ in (d). 
The overall numbers of dead under both scenarios are compared in (e).}
\label{fig:delta2}
\end{figure}

We end this two-scenario comparison by varying the ``high mortality rate'' $\delta_2,$ which is in effect when the number of Infected individuals exceeds a threshold ($I(t) \geq d_2 = 0.25$).
The results of these experiments are shown in Figure~\ref{fig:delta2}.
The color of curves in subfigures (a)-(d) indicated the value of $\delta_2$,
with the darkest blue showing the case with a constant/low mortality rate
($\delta(I) = \delta_2 = \delta_1 = 0.001$) and the next (light blue) curve 
corresponding to our default parameter value $\delta_2 = 0.002.$

\FloatBarrier

As $\delta_2$ grows, a traditional SIRSD model yields a much lower number of currently Infected individuals throughout the course of epidemic (subfigure (a)) since so many of them die (subfigure (e)) when $I(t)$ is sufficiently high.
(We recall that, based on \eqref{eq:death_rate}, the mortality rate starts to increase significantly 
from $\delta_1 = 0.001$ towards $\delta_2$ once $I(t)$ exceeds a lower threshold $d_1 = 0.15.$)
The peak of infection also decreases (as $\delta_2$ grows) in the Optimizers-only scenario 
(subfigure (b)).  But this is primarily driven by their sharp decrease in contact rates when $I(t)$ is high (subfigure (c)), leading to a decrease in the mortality rate to 
$\delta_1$
by the time $t=40$ even for the highest explored $\delta_2$ values (subfigure (d)).
As a result, the overall number of dead in this Optimizers-only scenario grows very slowly with $\delta_2;$ 
see Fig.~\ref{fig:delta2}(e).

\subsection{B-MFG: the impact of different behavioral patterns}

In this subsection we explore the impact of different behavioral patterns (subpopulations of Reckless, Followers, and Optimizers) and behavioral switches on the course of epidemics.  Focusing first on the simplified scenario of no switches
($\lambda_{j,k} = 0$ for all $j,k \in \{r,f,o\}$), we consider the basic interactions between the two subpopulations considered before: the Reckless and Optimizers (for now, assuming that $S_f(0) = I_f(0) = R_f(0) = 0$).  
The key parameter here will be $\rho_r \in [0,1]$, the initial fraction\footnote{
In all experiments without behavioral switching, the fractions of subpopulations can only change over time due to mortality differences resulting from their different contact rates.} of Reckless in the population.  Since the Followers are absent and we assume proportional representation among those 
initially infected, this means that $I_r(0) = \rho_r I(0), \, I_o(0) = (1-\rho_r) I(0),$ 
$S_r(0) = \rho_r S(0), \, S_o(0) = (1-\rho_r) S(0)$ with $R(0) = 0$ and $S(0) = 1 - I(0).$

\begin{figure}[H]
    \centering
    \includegraphics[height=4cm]{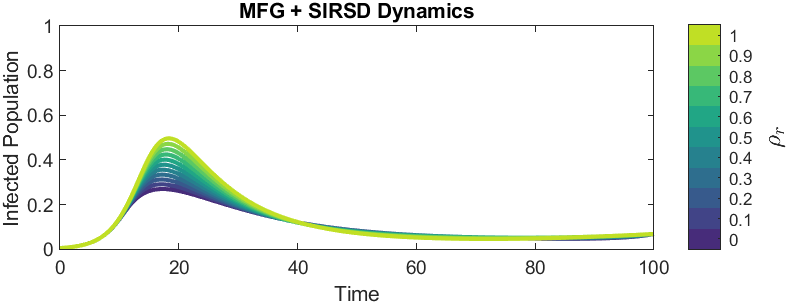}
    \hspace*{1.1mm}\includegraphics[height=4cm]{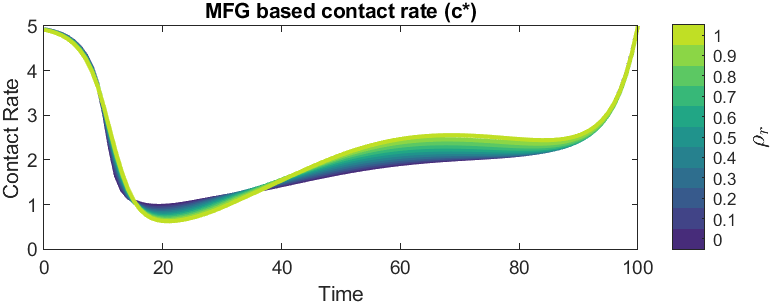}\hspace*{-1.1mm}
    \caption{Varying the initial share of Reckless individuals ($\rho_r \in [0,1]$) in a ``Reckless and Optimizers'' scenario without behavioral switches. (a) Total infected population $I(t)$. (b) Contact rate of Susceptible Optimizers, $c^*(t)$.  The colors of curves indicate the $\rho_r$ values.  The dark blue curves 
    correspond to a traditional MFG-SIRSD model -- a homogeneous all-Optimizers population ($\rho_r = 0$).  The \newest{bright yellow} $I(t)$ curve corresponds to the traditional SIRSD model -- a homogeneous all-Reckless population
    ($\rho_r = \newest{1}$), while the \newest{bright yellow} $c^*(t)$ curve shows the contact rate that a single Susceptible Optimizer would use (if added to this scenario, with no hope of affecting the course of epidemic).
    }
    \label{fig:R_percentage}
\end{figure}

\begin{figure}[hb]
    \centering
    $
    \begin{array}{cc}
    \includegraphics[height=5.5cm]{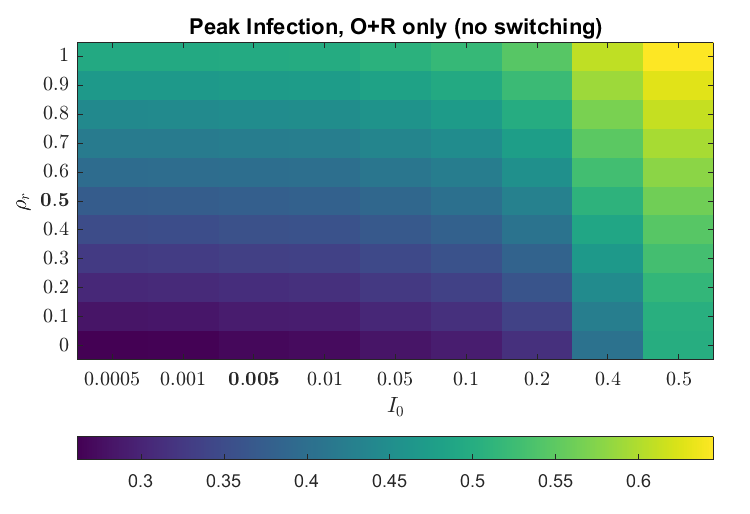}
    &
    \includegraphics[height=5.5cm]{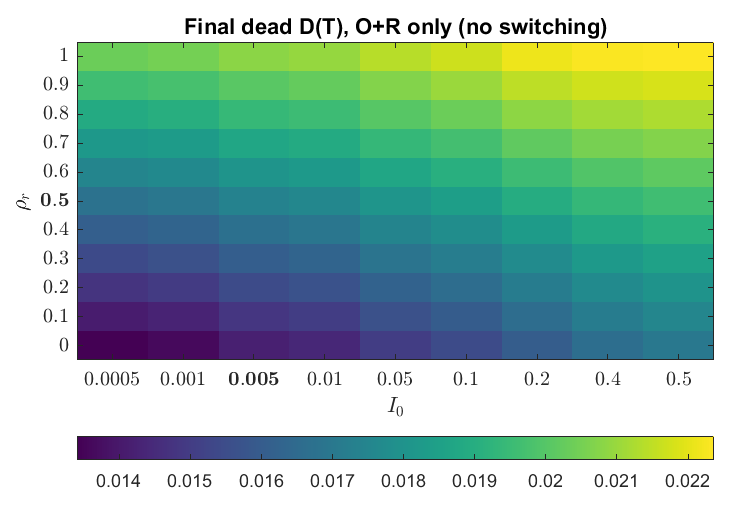}
    \end{array}$
    \caption{Epidemics outcomes varying in response to changes in parameters
$I_0$ (the initial fraction of Infected, spread proportionally between the two behavioral subpopulations: Reckless and Optimizers) and $\rho_r$ (the initial fraction of Reckless).
The peak of infection (defined as $\max_{t \in [0, T/2]} I(t)$) is shown on the left
while the total number of dead $D(T)$ is shown on the right.  
The axis are not uniformly scaled, with the sampled parameter values corresponding to epidemics trajectories shown in prior figures.  In particular, the horizontal lines $\rho_r = 0$ and $\rho_r = 1$ correspond to Fig.~\ref{fig:xi} while the vertical line 
$I_0 = 0.005$ corresponds to Fig.~\ref{fig:R_percentage}.
Both response variables exhibit monotone dependence on each of these two parameters.
}
    \label{fig:xi_rho_heatmap}
\end{figure}

As we show in Figure \ref{fig:R_percentage}, when the higher \newest{initial} proportion of the population is Reckless,
the total number of Infected individuals (and the peak of infection) become higher as well, despite the
Susceptible Optimizers' efforts. 
The effect on the total number of dead is similar.  These conclusions also hold true for other initial fractions of Infected; see Figure\ref{fig:xi_rho_heatmap}.
\newest{The higher $\rho_r$ 
leads Susceptible Optimizers to adopt lower contact rates around the time when infection peaks. But perhaps less intuitively, at later times their contact rates actually increase with $\rho_r.$ 
This is because a higher $\rho_r$ results in an increase in the number of Recovered individuals (after a higher peak of infection), which in turn increases the denominator in \eqref{eq:MFG_p_extended}, driving down the marginal contact risk.}

\begin{figure}[ht]
    \centering
    \begin{subfigure}{0.5\textwidth}
    \includegraphics[height = 0.8\linewidth]{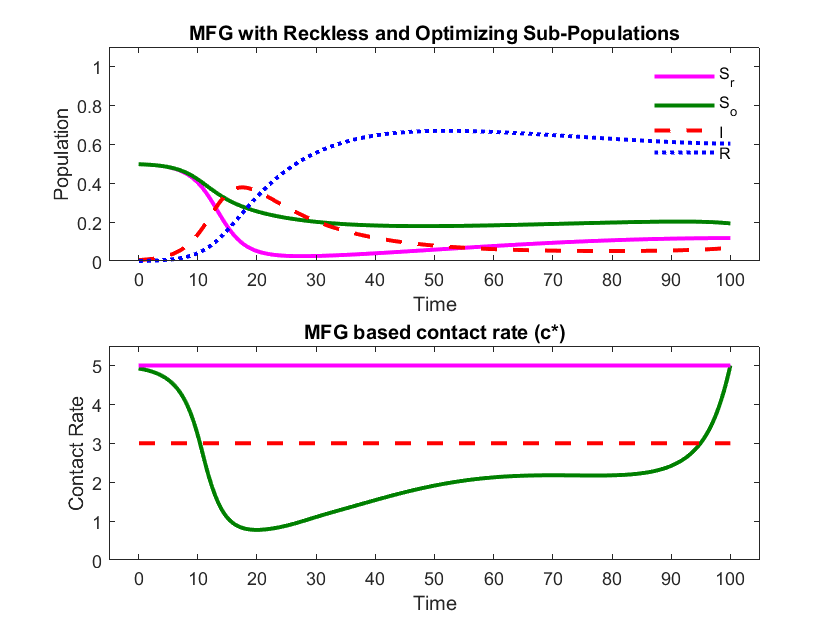}
    \caption{50\% Reckless, 50\% Optimizers}
    \end{subfigure}%
    \begin{subfigure}{0.5\textwidth}
    \includegraphics[height = 0.8\linewidth]{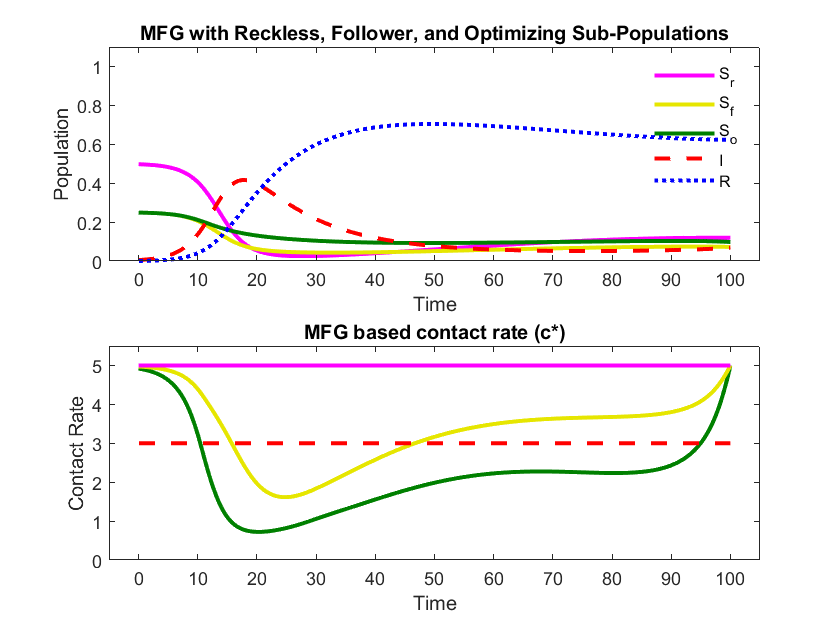}
    
    \caption{50\% Reckless, 25\% Optimizers, 25\% Followers}
    \end{subfigure}

    \caption{Two B-MFG scenarios to highlight the importance of 
    the initial fraction of Optimizers among the non-Reckless individuals (assuming no behavioral switches and that a half of population are initially Reckless). (a) All non-Reckless are Optimizers. (b) Non-Reckless are initially equally split between Optimizers and Followers. In the lower row, red indicates the contact rate of all Infected, while magenta, green, and yellow show the contact rates of Susceptible Reckless, Optimizers, and Followers respectively.  The peak of infection is $\approx\!\newest{37.8}\%$ in (a) and $\approx\!\newest{41.6}\%$ in (b).  The total \newest{percentage} of dead is $\approx\!\newest{1.7}\%$ in (a) and $\approx\!1.9\%$ in (b).      
    \label{fig:ROF_no_switching}
}
\end{figure}

We next consider the situation where $50\%$ of people are initially Reckless and 
compare the impact of Optimizers depending on their initial fraction:
either the remaining $50\%$ (as shown in Figure~\ref{fig:ROF_no_switching}(a))
or an even $25\%$ to $25\%$ split between the Optimizers and Followers 
(as shown in Figure~\ref{fig:ROF_no_switching}(b)).
Not surprisingly, the latter scenario results in a higher peak of infection and more deaths even though 
the Susceptible Optimizers adopt somewhat lower contact rates when half of them are replaced by Followers.
The Susceptible Followers' contact rates (shown in yellow) are qualitatively interesting.
They start fairly close to the Susceptible Reckless' rate $\cSbar$ (shown in magenta), since they are initially in the majority. But most of those Reckless become Infected quickly, so after $t \approx 18$
Susceptible Optimizers (whose contact rate is shown in green) are in the majority and Susceptible Followers mimic them much more closely.
This changes again as more Reckless individuals recover and start losing immunity, with the number of 
Susceptible Reckless and optimizers fairly comparable after $t \approx 60$ and Susceptible Followers 
equally influenced by them.

\FloatBarrier

\begin{figure}[ht]
    \centering
    \begin{subfigure}{0.5\textwidth}
    \includegraphics[height = 0.8\linewidth]{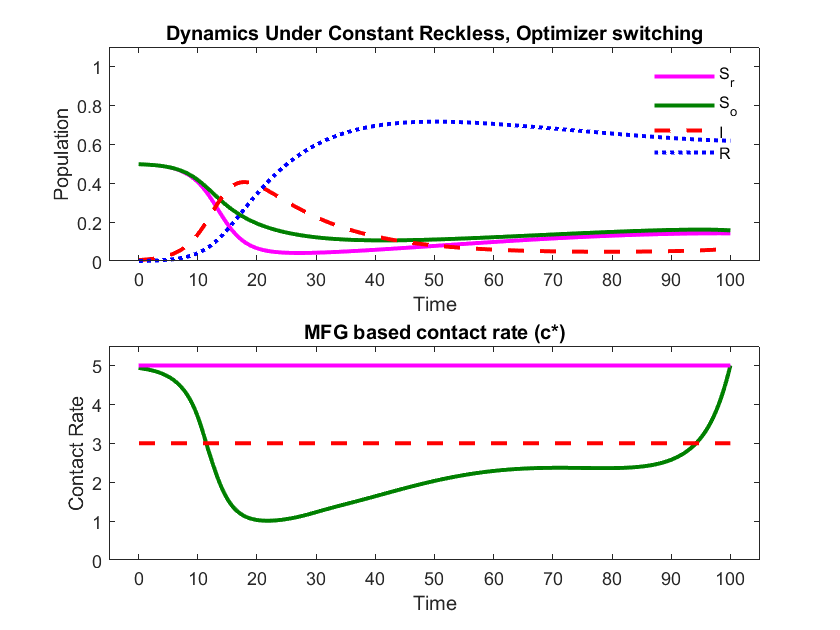}
    \caption{50\% Reckless + 50\% Optimizers to start,\\ $R \leftrightarrow O$ switching only, at rate $1/30$
    }
    \end{subfigure}%
    \begin{subfigure}{0.5\textwidth}
    \includegraphics[height = 0.8\linewidth]{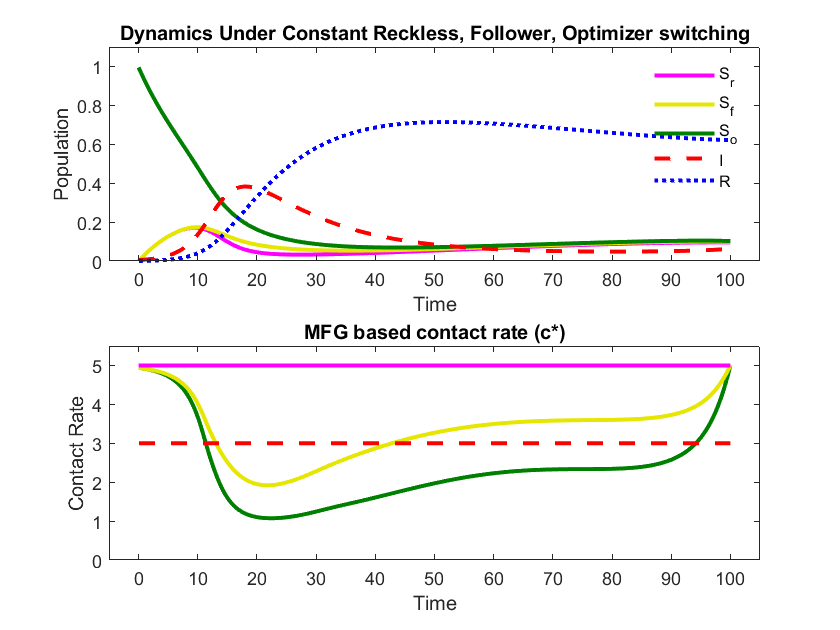}
    \caption{100\% Optimizers to start,\\ all switching rates $= 1/30$}
    \end{subfigure}
    \caption{Two scenarios with constant switching rates:
    (a) Reckless and Optimizers only, initially at $50\%$ each, with an equal rate of switching between them ($\lambda_{r,o} = \lambda_{o,r} = 1/30$) and no Followers; cf. Fig.~\ref{fig:ROF_no_switching}(a).  (b) Starting from Optimizers only, but with all six switching rates set to $\lambda = 1/30;$ cf. Fig.~\ref{fig:sirsd}(bc).}
    \label{fig:constant_switching}
\end{figure}

To illustrate the effect of behavioral switching, we first focus on constant switching rates and consider two different scenarios.
In the first (shown in Figure~\ref{fig:constant_switching}(a)), the initial population consists of an equal number of Reckless and Optimizers,
with an equal rate of switching between them ($\lambda_{r,o} = \lambda_{o,r} = 1/30$)
and no Followers ($\lambda_{r,f} = \lambda_{o,f} = 0$).
This is a switching-added modification of the example already presented in
Figure~\ref{fig:ROF_no_switching}(a).
In the second scenario (illustrated in Figure~\ref{fig:constant_switching}(b)), the initial population consists of Optimizers only, while Reckless and Followers emerge later as a result of symmetric behavioral switching ($\lambda_{j,k} = 1/30$ for all
$j \in \{r,f,o\}, \; k \in \left\{r,f,o\right\} \backslash \left\{ j \right\}$).
This is a switching-added modification of our very first MFG example in
Figure~\ref{fig:sirsd}.
Here, Subpopulations of Reckless and Followers grow quickly due to this behavioral switching.
Followers start by imitating Optimizers (who comprise most of the population for small $t$), but later use the contact rate much closer to 
$\left(\cSoS(t) + \cSbar \right)/2.$
In both scenarios,  
behavioral switches make it harder for the Optimizers to lower the peak of infection.
Indeed, the maximum fraction of simultaneously Infected is 
$\approx\!\newest{40.6}\%$ in Fig.~\ref{fig:constant_switching}(a)
compared to 
$\approx\!\newest{37.8}\%$ in Fig.~\ref{fig:ROF_no_switching}(a);
and also
$\approx\!\newest{38.3}\%$ in Fig.~\ref{fig:constant_switching}(b)
compared to 
$\approx\!\newest{27}\%$ in Fig.~\ref{fig:sirsd}(b).

\begin{figure}[ht]
\centering
\setcounter{subfigure}{0}

\begin{subfigure}{\textwidth}
  \centering
  \begin{minipage}{0.5cm}
    \refstepcounter{subfigure}%
    \makebox[\linewidth][l]{(\thesubfigure)} 
  \end{minipage}
  \begin{minipage}{12cm}
    \includegraphics[height=4cm]{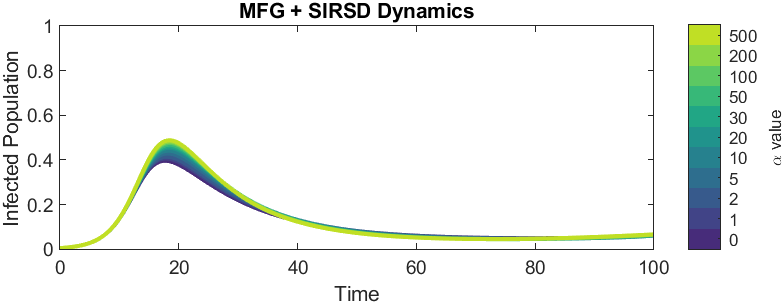}
  \end{minipage}\hspace{0.3em}%
  
\end{subfigure}

\begin{subfigure}{\textwidth}
  \centering
  \begin{minipage}{0.5cm}
    \refstepcounter{subfigure}%
    \makebox[\linewidth][l]{(\thesubfigure)}
  \end{minipage}
  \begin{minipage}{12cm}
  
  \hspace{0.1cm}
    \includegraphics[height=4cm]{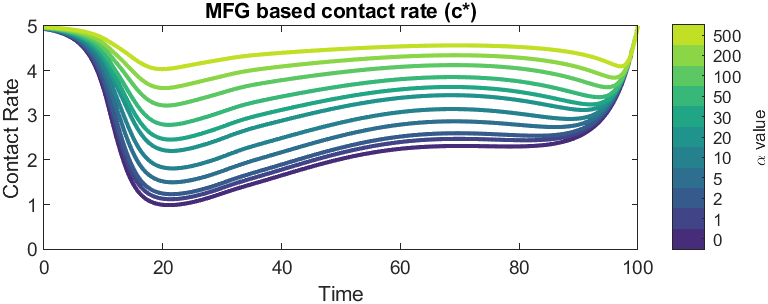}
  \end{minipage}\hspace{0.3em}%
  
\end{subfigure}

\hspace{0.03cm}
\begin{subfigure}{\textwidth}
  \centering
  \begin{minipage}{0.5cm}
    \refstepcounter{subfigure}%
    \makebox[\linewidth][l]{(\thesubfigure)}
  \end{minipage}
  \begin{minipage}{12cm}
    \includegraphics[height=4cm]{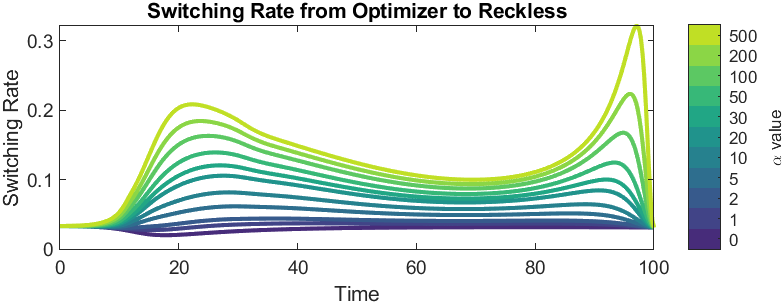}
  \end{minipage}\hspace{0.3em}%
  
\end{subfigure}

\caption{ Varying the \newest{coefficient} $\alpha$ in the ``Optimizers to Reckless'' switching rate function.
(a) Infected population, (b) optimal contact rates, and (c) $O \rightarrow R$ switching rates all shown as a function of time for various $\alpha$ starting from an initial population that consists of 50\% Reckless, 50\% Optimizers individuals.}
\label{fig:non_constant_switching_2}
\end{figure}

For our final example presented in Figure~\ref{fig:non_constant_switching_2}, we return to the 2-sub-populations setting (Reckless and Optimizers only) and explore the effect of non-constant behavioral switching by Optimizers.
In particular, we use
\begin{equation}
\label{eq:or_switching_rate}
    \lambda_{o,r} \; = \; 
    \frac{1}{30}\left(1-I(t)\right) \left(1 + \alpha \left[\uSbar - \uS\left(  \cSoS(t) \right) \right]\right),
\end{equation}
where $\alpha \geq 0.$  The factor $(1-I)$ models their lowered likelihood of switching to Reckless behavior when the proportion of infected people is high.
The last factor models an increased likelihood of becoming Reckless due to 
fatigue/frustration (since the chosen Nash-optimal contact rate lowers one's instantaneous utility).  To simplify the interpretation, we keep the rate 
of the reverse (``$R \rightarrow O$'') switching constant with $\lambda_{r,o} = \frac{1}{30}.$

\begin{remark}
Formula \eqref{eq:or_switching_rate} is a phenomenological model of endogenous behavioral switching rates, similar in spirit to several prior models reviewed in \cite{lejeune2025formulating}.  We chose it primarily as a simple illustration of qualitative features due to such $(I,\cSoS)$-dependent switching.
EGT-based models (described in Appendix \ref{appendix:other_games}) provide an alternative (principled and mechanistic) approach for deriving 
behavioral switch rates.  But we do not use it here for the sake of simplicity and to avoid the need to reconcile the MFG assumptions about long-term strategic planning with the social learning/imitation that drive EGT models.
\end{remark}

When $\alpha=0$, the case shown by the lowest (dark blue) curves in Figure~\ref{fig:non_constant_switching_2}, the fatigue/frustration has no influence, and ``$O \rightarrow R$'' switching is only decreased when the Infection is high.
This decrease makes it slightly easier for the Optimizers to lower the peak
of infection compared to Figure~\ref{fig:constant_switching}(a).
But for $\alpha>0,$ the Susceptible Optimizers face an interesting dilemma:
a lower contact rate decreases the probability of getting Infected in the near term, but increases the chances of switching to Reckless behavior (and hence also the chances of getting Infected in the future).  As shown in Fig.~\ref{fig:non_constant_switching_2}(b), higher $\alpha$ values force them to adopt higher contact rates, which leads to a higher proportion of Infected for $t \in [15, 35]$; see Fig.~\ref{fig:non_constant_switching_2}(a).
Nevertheless, they still use a noticeably lower contact rate near the peak of Infection.
For higher $\alpha$ values, there is also a significant decrease of contact rates near the end of the planning interval.  Even though this results in a spike in ``$O \rightarrow R$'' switching shown in Fig.~\ref{fig:non_constant_switching_2}(c),
the cure awaited by the time $T$ makes this strategy advantageous -- in those few remaining days, Optimizers would still prefer not to become Infected but are far less concerned with becoming Reckless.

\section{Discussion}
\label{s:conclusions}
We have presented a behavioral-epidemiological B-MFG model, in which 
a part of the population (the Optimizers) are ``highly rational'' in choosing their contact rate
while others either stick to their usual life pattern (the Reckless/stubborn) or 
are highly influenced by others (the Followers/consensus-seekers), 
adopting a contact rate averaged among all Susceptibles.
Importantly, each Optimizer makes their choice keeping in mind the behavior of other (non-rational) sub-populations,
the game-theoretic aspects (i.e., all Optimizers are making that choice simultaneously and thus affecting each other),
the human propensity to switch occasionally to different behavioral patterns, and the effects on the future epidemics trajectory of all of the above.  Their optimal (Nash-equilibrium) choice is modeled in the framework that extends the more usual approach of multi-population MFGs.  Since there are finitely many (health, behavioral) states,
this model yields a two point boundary value problem for a system of ODEs, which is readily solvable by standard numerical methods.

Our results in \S~\ref{s:results} confirm that selfish/independent decision makers can significantly reduce the peak infection count.  This is consistent with prior findings in Optimizers-only MFG models \cite{cho2020, aurell2022finite, aurell2022optimal}.  
This ``flattening of the curve'' also helps to reduce the number of disease-induced deaths, 
especially so if the mortality rate increases sharply when the number of Infected individuals exceeds some threshold. 
We show that the presence of other (``non-rational'') sub-populations can significantly limit the ability of Optimizers to produce this positive effect, even when they stick to much lower contact rates.
Qualitatively, this fact is not surprising (particularly when there are many Reckless individuals), but our model allows to quantify this difference. 
We also find that the impact of Followers depends on the relative abundance of Optimizers versus Reckless among those currently Susceptible.  Near the peak of the epidemic, when the higher percentage of Reckless are Infected or recently Recovered, the Susceptible Followers tend to be heavily influenced by Optimizers and thus more careful.  But this is partly due to our modeling assumption that their reference group are only 
those currently Susceptible. 

Focusing on modeling and computational aspects of B-MFGs, we have sidestepped important theoretical questions about the well-posedness of the 
resulting TPBVP problem (\ref{eq:MFG_ODEs_extended_S}-\ref{eq:MFG_ODEs_extended_VR}).  Based on our extensive numerical experiments, there always appears to be a unique Nash equilibrium, but it remains necessary to prove this, perhaps using a Banach fixed-point argument, similarly to how this was shown for traditional MFG models in \cite{aurell2022optimal, liu2025incorporating}.

There are several obvious extensions that would be relatively easy to implement
in our framework to broaden its applicability.  These include more sub-population types with different behavioral logic (e.g., semi-Reckless, who stick to a specific reduced contact rate level), effects of vaccinations, and different (e.g., health-state-dependent or peer-pressure-induced) behavioral switching rates. An extension to ``graphon-structure'' MFGs \cite{aurell2022finite} would also be natural to model the effects of travel restrictions 
and the age/frailty spectrum in the population. 
It will be harder but certainly 
useful
to include a refined model of fatigue-induced behavioral switches. Our simplified version presented here (based on the instantaneous utility reduction) ignores the build up of fatigue over time. In that sense, we model more of a current-life-style-frustration than adherence fatigue here. 

Some of the extensions that are relatively simple in standard epidemiological models become more subtle in the MFG framework.  E.g., adding Exposed, Presymptomatic, or Asymptomatic compartments means that some of the transitions in health status are not directly observable, and a rational Optimizer thus has to engage in ``belief space dynamic programming'' to determine their Nash-equilibrium contact rates
\cite{olmez2022modeling}.  Similar complications and numerical challenges arise in MFG models with an uncertain immunity \newest{status} \cite{doebeli2026partial},
leading to forward-backward integro-PDE systems. 
Nevertheless, these types of uncertainty, as well as uncertainty on the time of vaccine availability \cite{doebeli2026partial}, can be also implemented in our current B-MFG framework.
It would also be very useful to consider the effects of other uncertainty types (e.g., on vaccine efficacy or emergence of new viral strains) and partial or delayed information (e.g., about the current number of Infecteds).
Mean field control \cite{cho2020} and Stackelberg game \cite{aurell2022optimal} versions will also be of interest for such behaviorally-heterogeneous populations.  Another important next step will be to verify that
such behaviorally-enriched MFG models are better at matching the real data from past epidemics. 

While it is outside of our current scope, it will be important to conduct a series of careful head-to-head comparisons of MFG/B-MFG models with other game-theoretic approaches (including those described in Appendix \ref{appendix:other_games}) and explore the usefulness of hybrid versions.
On the fundamental level, MFGs make an important contribution to epidemiological literature by bringing rational choice of individuals to bear on the epidemic trajectory.  But prior MFG models take this to extremes by assuming that {\em everyone} is fully informed, perfectly rational, capable of long-term foresight, and consistent in carrying out their plans.  Here, we relaxed several of these limitations, but our assumptions about the predictive/cognitive abilities and self-awareness of Optimizers are still rather idealized.  It would be useful to develop models of ``time-limited'' 
predictive ability or predictive ability based on limited/local information, perhaps in the spirit of what was already proposed in MFG-models of crowd dynamics in traffic engineering \cite{cristiani2015modeling, carrillo2016improved}.  
Such hybrid models could incorporate the best features of MFGs and the ``stroboscopic planning'' approach \cite{fenichel2011adaptive} and would similarly benefit from considering behavioral patterns.  EGT-type mechanisms might also be employed to model the rate of behavioral switching in B-MFGs.  But the latter extension will require grappling with compatibility of imitation/social learning with long-term strategic planning.

To those used to rational decision theory, it might be tempting to seek a rational explanation for the behavior of each subpopulation; e.g.,  
a question about different utility functions that could rationalize the behavior of Reckless or Follower individuals could be viewed as an inverse problem for multi-population MFGs \cite{ren2025unique}. But we believe that such rationalizations would have limited practical impact in epidemiological context, and that it is more useful to admit that some (and perhaps most) humans are 
not making decisions based on game-theoretic considerations.
This is already well-recognized in epidemiological literature, particularly in 
papers on 
individual vaccination decisions.
While the classical game-theoretic models 
typically predict that herd immunity is unachievable without strong governmental policy
\cite{geoffard1997disease, bauch2004vaccination}, there is also  empirical evidence that vaccination decisions are often not ``strategic'' and that, at least with seasonal influenza vaccinations, individuals often disregard others' decisions when making their own \cite{parker2013conscious}.
This is why more recent work also includes a number of models 
where choices are based
on cognitive biases, various rules of thumb, and inertia \cite{voinson2015beyond, oraby2015bounded, papst2022modeling}.  Similarly, it is clear that non-rational 
decision making also plays a role in social distancing decisions and thus should be carefully considered 
in contact rate modeling.  

Finally, we must acknowledge that the 
model presented here remains primarily qualitative in nature.
To build  
accurate
predictive models that would be suitable
for evaluating policy proposals, 
descriptions of non-rational behavioral patterns and the switching rate functions should be developed in a close collaboration with domain experts: epidemiologists, social psychologists, and behavioral economists.
More generally, we hope that incorporating behavioral patterns and collaborating with domain experts will also improve the predictive power of MFGs as generative models of human behavior in other critical scenarios with complex social interactions (e.g., evacuating a building in case of emergency, retail investors reacting to major market news, dynamics of labor markets in the face of disruptive technologies, etc).


\addtolength{\textheight}{-0.0cm}


\vspace*{-3mm}

\section*{Acknowledgments}

This research was partially funded by the National Science Foundation (award DMS-2111522) 
as well as the Air Force Office of Scientific Research (award FA9550-22-1-0528)
and the second author's Royal Society Wolfson Visiting Fellowship.
A part of this work was done during the second author's sabbatical 
year at Imperial College London (ICL), and he is thus grateful to ICL for its hospitality.
The authors also thank A.~Gumel, S.~Ellner, and two anonymous reviewers for
pointing out many relevant papers and for providing very useful suggestions on improving this manuscript.


\vspace*{-3mm}

\section*{Data and Code Availability}

All data and code used to generate the figures in this manuscript are available online at\\ \url{https://github.com/eikonal-equation/Behavioral-MFGs-in-epidemiology}.

\begin{appendices}

\renewcommand{\thesection}{\Alph{section}}
\renewcommand{\appendixname}{Appendix}

\makeatletter
\renewcommand{\@seccntformat}[1]{%
  \appendixname\ \thesection.\ }
\makeatother

\section{Comparison of game-theoretic models in epidemiology}
\label{appendix:other_games}
The fundamental principles behind MFGs are mathematically elegant and logically appealing, which is probably why they have been 
independently (re)-invented by scholars working in several distinct areas.
Our basic (``everyone is an optimizer'') model described 
in \S~\ref{ss:Nash}
is close in spirit to others commonly presented in MFG papers; e.g., 
\cite{reluga2010game, cho2020, tembine2020covid, doncel2022mean, aurell2022optimal, aurell2022finite, olmez2022modeling, roy2023recent, bremaud2024mfg, bremaud2025mean, schnyder2025understanding, liu2025incorporating, doebeli2026partial}.
To the best of our knowledge, the first use of a similar approach in epidemiological literature is due to T.C. Reluga, developed for an SIR model with a density-based force of infection in 2010 \cite{reluga2010game}.  
Remarkably,
it appears that this model was derived without any knowledge of the general MFG framework, which was introduced four years earlier in two separate research communities on continuous state spaces \cite{lasry2006jeux_1,lasry2006jeux_2,huang2006large}.  Moreover, Reluga's work also predates the first papers proving the well-posedness of general continuous time / finite state MFG models \cite{gomes2013continuous}, which are the most relevant for the current context.

But MFGs are certainly not ``the only game in town''.
In this appendix we briefly review the fundamental modeling assumptions in two other game-theoretic approaches used in epidemiological modeling, with their key features summarized in Table \ref{tab:game_models}.

\begin{table}[h]
\footnotesize 
\begin{tabular}{||l || c | c | c | c||} 
 \hline
   & EGT-based & ``Stroboscopic'' & MFG & B-MFG \\ [0.5ex] 
 \hline\hline
Representative papers: 
& \cite{glaubitz2020oscillatory, martcheva2021effects, agusto2022isolate, saadroy2023dynamics}
& \cite{fenichel2011adaptive, espinoza2022heterogeneous, espinoza2024adaptive} 
& \cite{reluga2010game, aurell2022optimal, liu2025incorporating, schnyder2025understanding}
& this paper\\
\hline
Strategic planning? & no & yes & yes & yes\\
\hline
\multirow{2}*{Planning mode?} & 
\multirow{2}*{none} & 
frequent/iterative & only once & only once\\ [-1ex]
& & (rolling horizon) & (fixed horizon) & (fixed horizon)\\
\hline
Predictive ability? & no & no & yes & yes\\
\hline
Behavioral Patterns? & yes & no & usually none & yes \\
\hline
Behavioral Switching? & yes & no & no & yes \\
\hline
Social learning? & yes & no & no & no \\
\hline
Reaction to unforeseen changes? & automatic & automatic & with re-planning & with re-planning \\
\hline
Computational cost? & low & the highest & high & even higher \\  
 \hline
\end{tabular}
\caption{\label{tab:game_models}
A \newest{condensed}
summary of game-theoretic modeling approaches in epidemiology.  
}
\end{table}

One of these alternatives was first developed in a paper by Fenichel et al. \cite{fenichel2011adaptive} in 2011
and greatly extended in a sequence of more recent papers
\cite{fenichel2013economic, morin2013sir, perrings2014merging, espinoza2022heterogeneous, espinoza2024adaptive, espinoza2025impact}.  Using the same instantaneous utility functions that we have employed in section \S~\ref{ss:utility}, that paper 
modeled how susceptible individuals choose their contact rates in an SIR epidemic with a frequency-based  force of infection.  But despite these similarities, there are two key distinctions from MFGs, which have important practical, mathematical, and computational consequences.  First, in \cite{fenichel2011adaptive} it is assumed that all individuals have a shorter planning horizon $\Theta < T$ used for determining their ``optimal'' contact rates\footnote{
These quotation marks are meant to
emphasize the differences in optimization goals, optimization horizons, and available information across the spectrum of game-theoretic models.  A contact rate policy $\cSS$ Nash-optimal under MFGs is generally not optimal in the setting of \cite{fenichel2011adaptive} (and vice versa).
} 
$\cSmyo$
by solving the backwards-in-time ODEs satisfied by value functions on the time interval $[t, t+ \Theta].$  The terminal conditions for those value functions are consistent with an individual belief that the epidemic ends at the time $t+\Theta,$ even if in reality it continues much longer.
This makes it necessary to perform frequent {\em rolling-horizon re-planning},
which we describe below.
Second, it is assumed that individuals do not try to predict (or are not capable of predicting) the future epidemic trajectory $\y(\nt)$ even over the next $\Theta$ time units. Instead, they use a current ``snapshot'' for future planning; i.e.,
all susceptible individuals assume that the current state of the epidemic $\y(t) = \left( S(t), I(t), R(t), D(t) \right)$ will remain in effect until the time $t+\Theta.$   To reflect these assumptions, we will refer to such decision-making models as ``stroboscopic re-planning''.

Of course, once all Susceptibles start to use those contact rates, the epidemiological situation changes, and model consistency requires re-optimizing the contact rates over a new (shifted) time interval.  
Since $\y$ evolves in continuous time, it would seem natural to carry out this re-planning continuously as well.  I.e., for each time $t$, defining the Susceptibles' {\em current} ``stroboscopically optimal'' contact rate $\cSmyo(t)$ requires solving an auxiliary ODE system for $\nt \in [t, t + \Theta],$ with value functions computed based on an assumption $\y(\nt) = \y(t).$  Aside from the theoretical challenges in proving the well-posedness, this interpretation would also make the use of such ``stroboscopic re-planning'' models very expensive computationally.  (In MFGs, a single TPBVP problem is posed over $[0,T].$ Here, it is replaced by one initial value problem for $\y(t)$ over $[0, T]$ coupled with infinitely many terminal value problems (one for each $t \in [0,T]$), each posed over a different time interval of duration $\Theta$.)   
Practical implementations in \cite{fenichel2011adaptive} and subsequent papers reduce this computational burden by using instead a discrete-in-time re-planning process,
which might be also more realistic. 
Namely, it is assumed that $\Theta = n \theta$ for some integer $n$ and a sufficiently small $\theta$.  The ``stroboscopically optimal''  $\cSmyo$ found over the interval $[t, t + n \theta]$ is then used by Susceptibles not just at the time $t$ but for all $\nt \in [t, t+ \theta],$
with the next re-planning occurring only at the time $(t+ \theta)$ over the interval $[t+ \theta, t + (n+1) \theta]$
and the updated $\cSmyo$ then used for all $\nt \in [t + \theta, t + 2\theta],$ etc.
The advantages of such stroboscopic / rolling-horizon planning include (a) avoiding unrealistic assumptions about very strong predictive abilities of individuals and (b) a decision making process that does not require knowing the overall simulation duration $T.$
The stroboscopic perspective also sidesteps all issues related to Nash equilibria (because during $[t, t+\theta]$ all agents act as if they do not impact each other), and frequent re-planning automatically handles any unforeseen changes in epidemic trajectory.
In addition to its high computational cost, the drawbacks of this approach include strong model dependence on specific values of parameters $\Theta$ and $\theta$ as well as the need to define and motivate the terminal conditions imposed at the end of each planning horizon.
Just as in MFGs, this approach does not account for different behavioral patterns unrelated to health status changes; incorporating them (and behavioral switching) would be similarly useful.

\begin{remark}
It is worth noting that this general ``stroboscopic optimization'' approach (based on continuous re-planning, assuming at every instant that the current state of population will not change in the future) was also independently developed in another research community even earlier than the MFG framework.
In transportation engineering, this was the main idea in Hughes' seminal paper on how pedestrians plan their individual paths through a crowd \cite{hughes2002}.
This 2002 paper was followed by a flurry of research activity, including attempts to put it on firm theoretical ground \cite{difrancesco2011}, make it computationally efficient \cite{huang2009revisiting}, extend it to multiple interacting sub-populations of decision makers \cite{jiang2009reactive, cartee2017anisotropic}, and create hybrid versions with MFGs \cite{cristiani2015modeling}.  Some of the ideas in the latter paper could also be useful in epidemiological context to explore more realistic middle ground between the extremes of MFGs (full predictive ability over a long horizon $[t,T]$) and stroboscopic re-planning (no ability to predict any future changes in $\y$ at all). 
While such (short term prediction) hybrid models remain outside the scope of our current paper, we hope that they will be explored in the future. 
\end{remark}

A very different game-theoretic modeling approach is based on Evolutionary Game Theory (EGT)  \cite{smith1982evolution}.
EGT was first introduced as a framework for modeling proliferation of inherited traits in evolving biological populations
but later found many additional applications in modeling imitation and social learning among humans.
The number of epidemiological models of this type is high and we cite only a few representative examples: \cite{manfredi2013modeling, glaubitz2020oscillatory, martcheva2021effects, agusto2022isolate, saadroy2023dynamics, agusto2026algorithms}.  In a typical setting, people are split into a small number of behavioral patterns or strategy groups (e.g., those who engage in social distancing and those who do not), with constant contact rates pre-defined for each group.  The usual health status ODEs (e.g., based on an SIR model) are augmented in the spirit of a ``Replicator equation'' \cite{taylor1978evolutionary} to reflect how people switch among behavioral groups as the epidemiological situation + the popularity of different behavioral patterns (and thus also the current ``payoff'' of each of them) change with time.  
The switches are assumed to happen when people compare their recent payoffs, and the likelihood of each switch is typically proportional to the perceived advantage of other's strategy.  Since there is no optimization over any time horizon, the participants are not viewed as strategic planners or informed decision makers; instead, they switch their behavior reactively, striving to learn from others about strategies that become more successful in response to changing circumstances.  The advantages include relative simplicity and low computational cost of the resulting ODE models, with classical tools of dynamical systems theory directly usable to analyze the asymptotic behavior. 
The lack of strategic planning also ensures that the frequency of behavioral patterns automatically adjusts in response to any unforeseen changes.

\section{Asymptotic behavior of SIRS model.}
\label{appendix:SIRS_equilibria}
With $\delta_1=\delta_2=0,$ our baseline model from \S~\ref{ss:baseline_model} becomes equivalent to an SIRS with the frequency-based force of infection, proportionate mixing, and health-status-dependent constant contact rates $(\cS,\cI,\cR)$.  
While models with proportionate mixing are well-studied \cite{nold1980heterogeneity, HETHCOTE198785}, 
the asymptotic analysis for this specific variation does not appear to be covered in the literature, and we include it here for the sake of completeness.

At the start of an epidemic, $S \approx 1$ and, from formula \eqref{eq:original_p}, the probability of each Susceptible becoming Infected per contact
is $p \approx \beta  \cI I/ \cS.$
The basic reproductive number is thus $R_0=\frac{\beta \cI}{\mu}.$
If the contact rates are the same across all
health states,  the following result reduces to a well-known property
of the classical SIRS model.  But in our heterogeneous case,
it is interesting that the stability of the endemic equilibrium does not depend on exact positive values of $\cS$ and $\cR$.
\begin{prop}
Suppose the SIRS model is described by ODEs (\ref{eq:SIRSD_ODEs_original_S}-\ref{eq:SIRSD_ODEs_original_R}) with $\delta(I) \equiv 0$.
Suppose also that the parameters $\beta, \mu, \gamma$ and all contact rates $(\cS,\cI,\cR)$ are positive. 
When $R_0 < 1,$ the system does not have an endemic equilibrium 
and the disease-free equilibrium $(S=1, I=0, R=0)$ is locally asymptotically stable. When $R_0 > 1,$ the disease-free equilibrium becomes unstable and there exist an endemic equilibrium $(S^*, I^*, R^*),$ which is locally asymptotically stable with no additional assumptions about $(\cS,\cI,\cR).$
\end{prop}

\begin{proof}
With the  number of people conserved, $S=1-I-R$
reduces the system to 2D:
\begin{equation}
\label{eq:2D_reduced}
\frac{dI}{dt}=\frac{\beta \cS\cI(1-I-R)I}{\cS(1-I-R)+\cI I+\cR R}-\mu I,
\hspace{1.75cm}
\frac{dR}{dt}=\mu I-\gamma R.    
\end{equation}
Supposing that $I\neq 0$ and setting $I'=R'=0$, we derive the 
formula for the equilibrium as 
\[I^* =\frac{\cS \gamma (R_0-1)}{\cI\gamma +\cR\mu+\cS(\gamma+\mu)(R_0-1)}\]
\[
R^*=\frac{\mu}{\gamma}I^*=\frac{\cS \mu (R_0-1)}{\cI\gamma +\cR\mu+\cS(\gamma+\mu)(R_0-1)},
\qquad
S^*=1-\frac{\gamma+\mu}{\gamma}I^*=
\frac{\cI\gamma +\cR\mu}{\cI\gamma +\cR\mu+\cS(\gamma+\mu)(R_0-1)}.
\]
Note that $R_0>1$ implies the existence of an endemic equilibrium (since 
$S^*, I^*, R^* > 0$), while $R_0=1$ is a transcritical bifurcation point, at which 
$(S^*, I^*, R^*) = (1,0,0),$ coincides with the disease-free equilibrium.
With $R_0 < 1,$ for $I^*$ to be positive, the other parameters would have to be such that the denominator is negative.  But that would result in $S^*<0.$
Thus, the endemic equilibrium exists if and only if $R_0 > 1.$

Next, we consider stability of the disease-free equilibrium where $I=R=0$.\\
Defining $Q(I,R)=\cS(1-I-R)+\cI I+\cR R$, we can write the Jacobian of \eqref{eq:2D_reduced} as
\begin{equation}
\label{eq:Jac}
J(I,R)=\begin{pmatrix}
\frac{\partial I'}{\partial I} &
\frac{\partial I'}{\partial R}\\
\frac{\partial R'}{\partial I} & 
\frac{\partial R'}{\partial R}
\end{pmatrix}=
\begin{pmatrix}
\mu\left[
R_0 \cS
\frac{(1-2I-R)Q-I (1-I-R) (\cI-\cS)}
{Q^2}
-1
\right]
&
-\mu R_0 \cS
\frac{IQ+I(1-I-R)( \cR- \cS)}
{Q^2}
\\[2.2ex]
\mu
&
-\gamma
\end{pmatrix}.
\end{equation}
At the disease-free equilibrium, this reduces to 
$J(0,0)=\begin{pmatrix}
\mu(R_0-1) & 0\\
\mu & 
-\gamma
\end{pmatrix}.$\\
Thus, $(S=1, I=0, R=0)$ is locally asymptotically stable for $R_0 < 1$
and unstable for $R_0>1.$

To analyze the stability of the endemic equilibrium,
we first note that $Q^* = Q(I^*,R^*) = R_0\cS S^* > 0.$ 
Using \eqref{eq:Jac} and simplifying, we obtain
\begin{equation*}
J^*
=
J(I^*,R^*)
=
\begin{pmatrix}
\displaystyle
-\mu I^*
\frac{\cI + \cS(R_0-1)}{Q^*}
&
\displaystyle
-\mu I^*
\frac{\cR + \cS(R_0-1)}{Q^*}
\\[3ex]
\mu
&
-\gamma
\end{pmatrix}.
\end{equation*}
When $R_0 > 1,$ we see that
$\operatorname{tr}(J^*)
=
-\gamma
-
\frac{\mu I^*}{Q^*}
\left[\cI+\cS(R_0-1)\right] < 0
$
 and 
$\det(J^*)
=
\frac{\mu\gamma \cS(R_0-1)}{Q^*} > 0,$
guaranteeing the local asymptotic stability of $(S^*,I^*,R^*).$
\end{proof}

\end{appendices}

\bibliographystyle{besjournals}
\bibliography{references}



\end{document}